\documentclass[11pt,a4paper]{article}
\usepackage[hyperref]{naaclhlt2019}
\usepackage{times}
\usepackage{latexsym}
\usepackage{ifthen}
\usepackage{dsfont}
\usepackage{url}

\usepackage{fdsymbol}
\def\aclpaperid{1096} %  Enter the acl Paper ID here

\usepackage{xcolor}
\usepackage{multirow}

\usepackage{mathtools}
\usepackage{amsmath}
\mathtoolsset{showonlyrefs}

\usepackage[ruled,norelsize]{algorithm2e}
\SetKwComment{Comment}{$\triangleright$\ }{}

\usepackage{tabulary}
\usepackage{graphicx}
\usepackage{enumerate}
\usepackage{subcaption}
\usepackage[normalem]{ulem}

\newcommand\refresh{\textsc{Refresh}}
\newcommand\ExConSumm{\textsc{ExConSumm}}
\newcommand\bs{\boldsymbol}
\definecolor{forestgreen}{HTML}{009B55}
\definecolor{sepia}{HTML}{671800}
\definecolor{midnightblue}{HTML}{006795}
\definecolor{orangered}{HTML}{ED135A}

\newcommand{\ST}[1]{\sout{\textcolor{purple}{#1}}}

\newcommand\blfootnote[1]{%
  \begingroup
  \renewcommand\thefootnote{}\footnote{#1}%
  \addtocounter{footnote}{-1}%
  \endgroup
}

\aclfinalcopy % Uncomment this line for the final submission

\title{Jointly Extracting and Compressing Documents\\ with Summary State Representations}

\author{
Afonso Mendes$^{\spadesuit}$ \quad Shashi Narayan$^{\diamondsuit\ast}$
\quad Sebasti\~{a}o Miranda$^{\spadesuit}$ \\
\textbf{
\quad Zita Marinho$^{\heartsuit\spadesuit}$
\quad Andr\'e F. T. Martins$^{\dagger\clubsuit}$ 
\quad Shay B. Cohen$^{\diamondsuit}$ }
 \smallskip \\
$^\spadesuit$Priberam Labs, Alameda D.~Afonso Henriques, 41, 2$^{\mathrm{o}}$, 1000-123 Lisboa, Portugal \\
$^\diamondsuit$School of Informatics, University of Edinburgh, Edinburgh EH8 9AB, UK \\
$^\heartsuit$ Instituto de Sistemas e Rob\'otica, Instituto Superior T\'{e}cnico, 1049-001 Lisboa, Portugal \\
$^\dagger$Instituto de Telecomunica\c{c}\~{o}es, Instituto Superior T\'{e}cnico, 1049-001 Lisboa, Portugal \\
$^{\clubsuit}$Unbabel Lda, Rua Visconde de Santar\'em, 67-B, 1000-286 Lisboa, Portugal
 \smallskip \\
 {
{\tt  amm@priberam.com}, {\tt shashi.narayan@gmail.com}, {\tt  ssm@priberam.com},}\\
{{\tt zam@priberam.com}, {\tt andre.martins@unbabel.com}, 
{\tt scohen@inf.ed.ac.uk}}
}

\date{}

\begin{document}
\maketitle
\begin{abstract}
  We present a new neural model for text summarization that first extracts sentences from a document and then compresses them.
 The proposed model offers a balance that sidesteps the difficulties in abstractive methods while generating more concise summaries than extractive methods. In addition, our model dynamically determines the length of the output summary based on the gold summaries it observes during training, and does not require length constraints typical to extractive summarization. The model achieves state-of-the-art results on the CNN/DailyMail and Newsroom datasets, improving over current extractive and abstractive methods. Human evaluations demonstrate that our model generates concise and informative summaries. We also make available a new dataset of oracle compressive summaries derived automatically from the CNN/DailyMail reference summaries.\footnote{Our dataset and code is available at \url{https://github.com/Priberam/exconsumm}.} 
\end{abstract}

\section{Introduction}
\label{sec:intro}
\noindent Text summarization is an important NLP problem with a wide range of applications in data-driven industries (e.g., news, health, and defense). Single document summarization---the task of generating a short summary of a document preserving its informative content  \cite{jones2007automatic}---has been a highly studied research topic in recent years \cite{Nallapati2016Abstractive,See2017Get,Fan2017Controllable,Pasunuru-multireward18}.%Cheng2016Neural,Nallapati2017SummaRuNNer,Narayan2018Ranking,Paulus2018Deep, asli-multiagent18  \andre{We can select only the 3-4 most relevant cites here to save space. The other citations are repeated below.}

\blfootnote{$^\ast$ Now at Google London.}
Modern approaches to single document summarization using neural network architectures have primarily focused on two strategies: \emph{extractive} and \emph{abstractive}. The former select a subset of the sentences to assemble a summary~\cite{Cheng2016Neural,Nallapati2017SummaRuNNer,narayan-sidenet18,Narayan2018Ranking}. The latter generates sentences that do not appear in the original document \cite{See2017Get,narayan18:xsum,Paulus2018Deep}.
Both methods suffer from significant drawbacks: extractive systems are wasteful since they cannot trim the original sentences to fit into the summary, and they lack a mechanism to ensure overall coherence. In contrast, abstractive systems require natural language generation and semantic representation, problems that are inherently harder to solve than just extracting sentences from the original document.

\begin{figure}[t!]

  \center{\fontsize{8.5}{6.2}\selectfont % \scriptsize \footnotesize
    \begin{tabular}{| p{7.2cm} |}
      \hline 
      \vspace{0.02cm}
      \textbf{(\ExConSumm\ Extractive)} \textbullet \hspace{0.1cm} (CNN) A top al Qaeda in the Arabian Peninsula leader--who a few years ago was in a U.S. detention facility--was among five killed in an airstrike in Yemen, the terror group said, showing the organization is vulnerable even as Yemen appears close to civil war. \\\textbullet \hspace{0.1cm} Ibrahim al-Rubaish died Monday night in what AQAP's media wing, Al-Malahem Media, called a ``crusader airstrike.'' \\
      \vspace{0.1pt}
      \textbf{(\ExConSumm\ Compressive)} \textbullet \hspace{0.1cm} \sout{\textcolor{purple}{(CNN)}} A top al Qaeda in the Arabian Peninsula leader--who a few years ago was in a U.S. detention facility--was among five killed in an airstrike in Yemen \sout{\textcolor{purple}{, the terror group said, showing the organization is}} \sout{\textcolor{purple}{vulnerable}} \sout{\textcolor{purple}{even as Yemen appears close to civil war}}. \\\textbullet \hspace{0.1cm} Ibrahim al-Rubaish died \sout{\textcolor{purple}{Monday night}} in what AQAP's media wing, Al-Malahem Media, called a ``crusader airstrike.'' \\ 
      
     \hline      
      
    \end{tabular}
    }
  \caption{ Summaries produced by our model. For illustration, the compressive summary shows the removed spans strike-through.}\label{fig:summaries-intro}
  \vspace{-0.5cm}
\end{figure}

In this paper, we present a novel architecture that attempts to mitigate the problems above via a middle ground, {\bf compressive summarization}~\cite{Martins2009NAACL}. Our model selects a set of sentences from the input document, and compresses them by removing unnecessary words, while keeping the summaries informative, concise and grammatical. We achieve this by dynamically modeling the generated summary using a Long Short Term Memory (LSTM; \citeauthor{Hochreiter1997Long}, \citeyear{Hochreiter1997Long}) to produce {\bf summary state representations}. This state provides crucial information to iteratively increment summaries based on previously extracted information. It also facilitates the generation of variable length summaries as opposed to fixed lengths, in previous extractive systems~\cite{Cheng2016Neural,Nallapati2017SummaRuNNer,Narayan2018Ranking,zhang2018neural}. Our model can be trained in both extractive (labeling sentences for extraction) or compressive (labeling words for extraction) settings. Figure~\ref{fig:summaries-intro} shows a summary example generated by our model. 

Our contributions in this paper are three-fold:
\begin{itemize}
    \item we present the first end-to-end neural architecture for EXtractive and COmpressive Neural SUMMarization (dubbed \ExConSumm, see \S\ref{sec:exconsum}),
    \item we validate this architecture on the CNN/DailyMail and the Newsroom datasets \cite{Hermann2015Teaching,newsroom_N181065}, showing that our model generates variable-length summaries which correlate well with gold summaries in length and are concise and informative (see \S\ref{sec:results}), and
    \item we provide a new CNN/DailyMail dataset annotated with automatic compressions for each sentence, and a set of compressed oracle summaries (see \S\ref{setup}). 
\end{itemize}
Experimental results show that when evaluated automatically, both the extractive and compressive variants of our model provide state-of-the-art results. Human evaluation further shows that our model is better than previous state-of-the-art systems at generating informative and concise summaries.

\section{Related Work}
\label{sec:relatedwork}

Recent work on neural summarization has mainly focused on sequence-to-sequence (seq2seq) architectures~\cite{Sutskever2014Sequence}, a formulation particularly suited and initially employed for abstractive summarization~\cite{Rush2015Neural}. However, state-of-the-art results have been achieved by RNN-based methods which are extractive. They select sentences based on an LSTM classifier that predicts a binary label for each sentence~\cite{Cheng2016Neural}, based on ranking using reinforcement learning~\cite{Narayan2018Ranking}, or even by training an extractive latent model~\cite{zhang2018neural}.
Other methods rely on an abstractive approach with strongly conditioned generation on the source document~\cite{See2017Get}. In fact, the best results for abstractive summarization have been achieved with models that are more extractive in nature than abstractive, since most of the words in the summary are copied from the document \cite{gehrmann2018bottom}. 

Due to the lack of training corpora, there is almost no work on neural architectures for compressive summarization.
% \shaycomment{what is the difference between compressive summarization and extractive-compressive summarization? please let me also know by email, so I can see if this paragraph makes sense.}. 
Most compressive summarization work has been applied to smaller datasets~\cite{Martins2009NAACL, Berg-Kirkpatrick2011Jointly, almeida2013fast}.
Other non-neural summarization systems apply this idea to select and compress the summary. \newcite{Dorr03} introduced a method to first extract the first sentence of a news article and then use linguistically-motivated heuristics to iteratively trim parts of it. \newcite{DurrettBK16} also learns a system that selects textual units to include in the summary 
and compresses them by deleting word spans guided by anaphoric constraints to improve coherence. 
Recently, \newcite{zhang2018neural} trained an abstractive sentence compression model using attention-based sequence-to-sequence architecture \cite{Rush2015Neural} to map a sentence in the document selected by the extractive model to a sentence in the summary. However, as the sentences in the document and in the summary are not aligned for compression, their compression model is significantly inferior to the extractive model.

\begin{figure*}[t!]
  \centering
    \includegraphics[width=0.9\textwidth, scale=0.8]{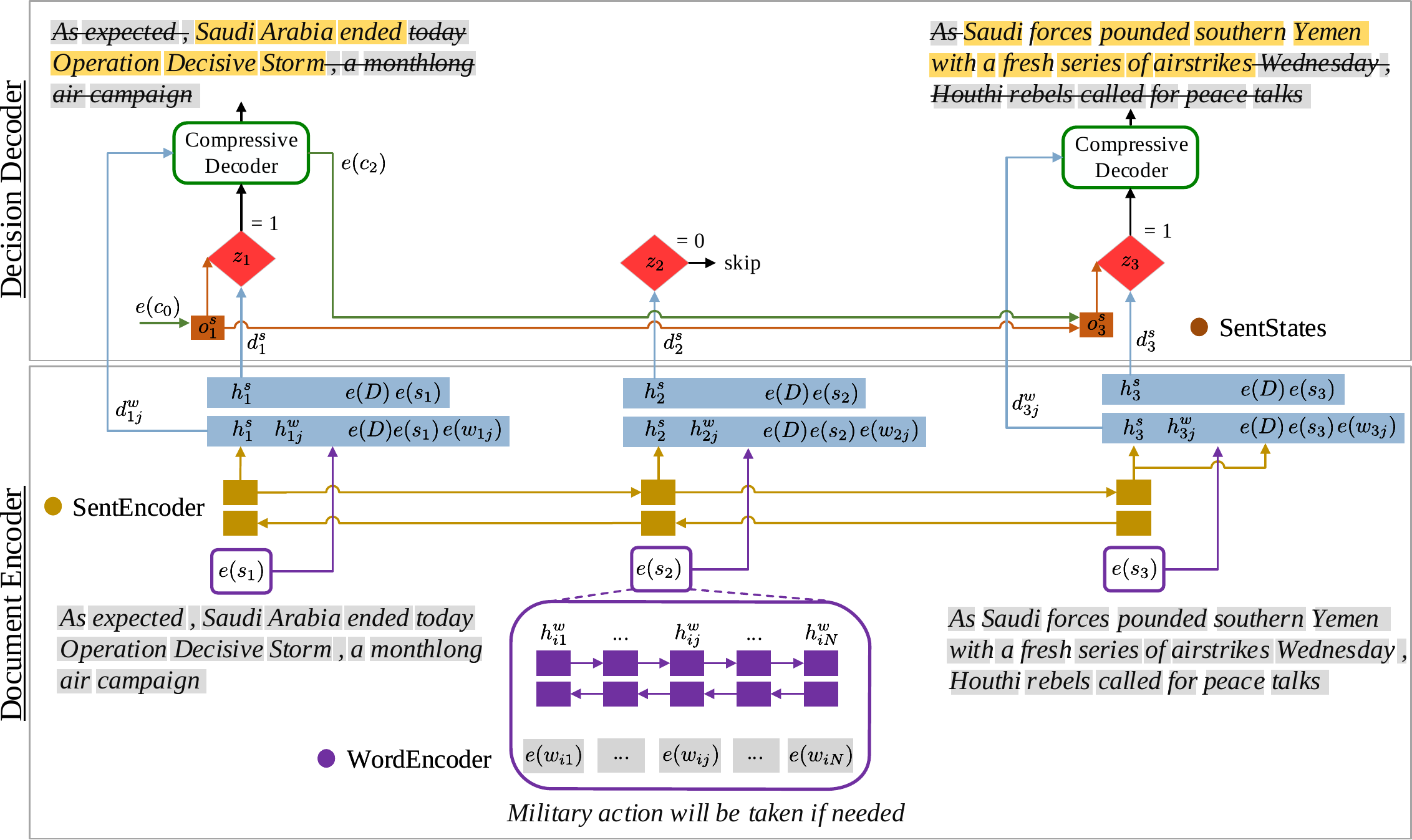}
    \caption{ Illustration of our summarization system. The model extracts the most relevant sentences from the document by taking into account the $\mathrm{WordEncoder}$ representation of the current sentence  $\bs e(s_i)$, the $\mathrm{SentEncoder}$ representation of the previous sentence $\bs h^s_{i}$, the current summary state representation  
    $\bs o^s_{i}$, and the representation of the document $\bs e(D)$. If a sentence is selected ($z_i=1$), its representation is fed to  $\mathrm{SentStates}$, and we move to the next sentence. Here, sentences $s_1$ and $s_3$ were selected. If the model is also compressing, the compressive layer selects words for the final summary (Compressive Decoder). 
    %At this level, the word representation is fed to the $\mathrm{WordStates}$ ($o^w_{ij}$) which conditions the selection of the next words. In addition to the same inputs, as described for the sentence extraction, we also add the $\mathrm{WordEncoder}$ state of the previous word $\bs h^w_{ij}$. 
    %The final output is read from the sentence-level decision labels ${z_i}$, for the extractive case, and from the word-level decision labels for the compressive case in (Compressive Decoder). 
    See Figure~\ref{fig:encdec} for details on the decoders.} %extractive and compressive decoders.}
    \label{fig:arquictecture}
\end{figure*}
In this paper, we propose a novel seq2seq architecture for compressive summarization and demonstrate that it avoids the over-extraction of existing extractive approaches~\cite{Cheng2016Neural, DlikmanLast16, NallapatiZM16}.
% In this paper, we propose a combination of a neural seq2seq architecture with an extractive and  compressive summarizer. Our objective is to combine a highly performant neural architecture with a compressive model, to avoid the over-extraction of existing extractive approaches~\cite{Cheng2016Neural, DlikmanLast16, NallapatiZM16}.

Our model builds on recent approaches to neural extractive summarization as a sequence labeling problem, where sentences in the document are labeled to specify whether or not they should be included in the summary~\cite{Cheng2016Neural, narayan-sidenet18}. These models often condition their labeling decisions on the document representation only. 
 % Other models explicitly model the summary being generated at every point. In this line, 
\newcite{Nallapati2017SummaRuNNer} tries to model the summary as the average representation of the positively labeled sentences. However, as we show later, this strategy is not the most adequate to ensure summary coherence, as it does not take the order of the selected sentences into account. 
Our approach addresses this problem by maintaining an LSTM cell to dynamically model the generated summary. 
To the best of our knowledge, our work is the first to use a model that keeps a state of already generated summary to effectively model variable-length summaries in an extractive setting, and the first to learn a compressive summarizer with an end-to end approach.

% Our model draws inspiration from the \textsc{StackLSTM} architecture proposed by \newcite{Lample2016Neural} for named-entity recognition. We adopt a summary state as a stack proxy to model the state of a finite state machine using a forward LSTM. However, instead of modeling the state of a transition based parser we adapt it for language generation based on the previous words and sentences already generated.

% To the best of our knowledge, our work is the first to use a model that keeps a summary of the already generated sentences and words in a compressive model for summarization.

\section{Summarization with Summary State Representation}
\label{sec:exconsum}
\begin{figure}[t]
\begin{subfigure}[b]{0.5\textwidth}
  \centering
    \includegraphics[trim={2.95in 0.5in 4.08in 1.5in}, clip,scale=0.76]{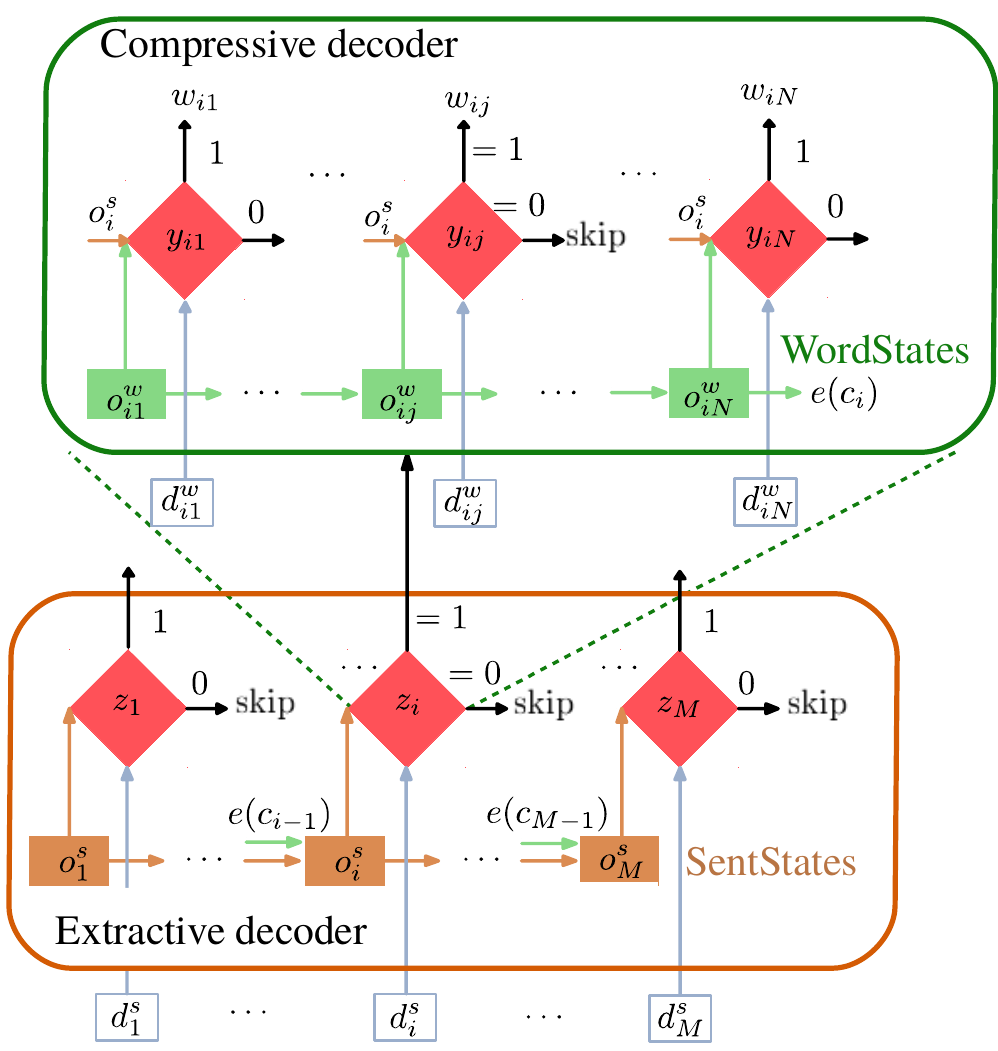}
\end{subfigure}
    \caption{ Decision decoder architecture. Decoder contains an extractive level for sentences (orange) and a compressive level for words (green), using an LSTM to model the summary state. Red diamond shapes represent decision variables $z_i=1$ if $p(z_i\mid \bs p_i)>0.5$ for selecting the sentence $s_i$, and $z_i=0$ if $p(z_i\mid \bs p_i)\leq0.5$ for skipping this sentence. The same for $y_{ij}$ and $p(y_{ij}\mid \bs q_{ij})>0.5$ for deciding over words $w_{ij}$ to keep in the summary.}
    \label{fig:encdec}
    \vspace{-0.2in}
\end{figure}

Our model extracts sentences from a given document and further compresses these sentences by deleting words. More formally, we denote a document $D = (s_1, \ldots, s_M)$ as a sequence of $M$ sentences, and a sentence $s_i = (w_{i1},\ldots,w_{iN})$ as a sequence of $N$ words. We denote by $\bs e(w_{ij})$, $\bs e(s_i)$ and $\bs e(D)$ the embedding of words, sentences and document in a continuous space.
We model document summarization as a sequence labeling problem where the labeler transitions between internal states. Each state is dynamically computed based on the context, and it combines an extractive summarizer followed by a compressive one. First, we encode a document in a multi-level approach, to extract the embeddings of words and sentences (``Document Encoder'').
Second, we decode these embeddings using a hierarchical ``Decision Decoder.'' The extractive summarizer labels each sentence $s_i$ with a label $z_i \in \{0, 1\}$ where~$1$ indicates that the sentence should be included in the final summary and $0$ otherwise. An extractive summary is then assembled by selecting all sentences with the label $1$. Analogously, the compressive summarizer labels each word $w_{ij}$ with a label $y_{ij} \in \{0, 1\}$, denoting whether the word $j$ in sentence $i$ is included in the summary or not. The final summary is then assembled as the sequence of words $w_{ij}$ for each $z_i = 1$ and $y_{ij} = 1$. See Figures~\ref{fig:arquictecture} and~\ref{fig:encdec} for an overview of our model. 
We next describe each of its components in more detail.

% \iffalse
% \begin{figure*}
%   \centering
%     \includegraphics[width=0.75\textwidth]{StackSumm11_zoom_1.pdf}
%     \caption{\footnotesize Illustration of the compressive layer of \ExConSumm processing an example sentence which has been selected by the model.}
%     \label{fig:encdec}
% \end{figure*}
% \fi
%\shaycomment{Afonso, I found a data structure called ``pop stack'' in which the pop is replaced with a pop that pops everything. it is used as a sorting device. in any case, I suggest somewhere we make it clear we use pop stacks and not regular stacks.}

\subsection{Document Encoder}
%As LSTMs are known to learn to extract information from arbitrary points in their memory \cite{Hochreiter1997Long}, they are ideal for modeling the document and the sentences.
The document encoder is a two layer biLSTM, one layer encoding each sentence, and the second layer encoding the document. The first layer takes as input the word embeddings $\bs e(w_{ij})$ for each word $j$ in sentence $s_i$, and outputs the hidden representation of each word $\bs h^w_{ij}$. The hidden representation consist of the concatenation of a forward $\overrightarrow{\bs{h}}^w_{ij}$ and a backward $\overleftarrow{\bs{h}}^w_{ij}$ LSTM (\textrm{WordEncoder} in Figure~\ref{fig:arquictecture}). This layer eventually outputs a representation for each sentence $\bs e(s_i)=[\overrightarrow{\bs{h}}^w_{iN},  \overleftarrow{\bs{h}}^w_{i1} ]$ that corresponds to the concatenation of the last forward and first backward LSTMs. The second layer encodes information about the document and is also a biLSTM that runs at the sentence-level. This biLSTM takes as input the sentence representation from the previous layer $\bs e(s_i)$ and outputs the hidden representation for each sentence $s_i$ in the document as $\bs h^s_i$ (\textrm{SentEncoder} in Figure~\ref{fig:arquictecture}). We consider the output of the last forward LSTM over M sentences and first backward LSTM to be the final representation of the document $\bs e(D)=[\overrightarrow{\bs{h}}^s_{M},  \overleftarrow{\bs{h}}^s_{1} ]$.

The encoder returns two output vectors, $\bs d^s_i=[\bs e(D),\bs e(s_i), \bs h^s_i]$ associated with each sentence $s_i$, and $\bs d^w_{ij}=[\bs e(D), \bs e(s_i), \bs e(w_{ij}), \bs h^s_i, \bs h^w_{ij}]$ for each word $j$ at the specific state of the encoder $i$.

\subsection{Decision Decoder}
Given that our model operates both at the sentence-level and at the word-level, the decision decoder maintains two state LSTMs 
%``pop stacks'',\footnote{A pop stack \cite{avis1981pop} operates as a normal stack, only the pop operation is replaced with one that empties the stack by popping {\em all} its elements.} 
denoted by $\mathrm{SentStates}$ and $\mathrm{WordStates}$ as in Figure~\ref{fig:encdec}. For the sentence-level decoder sentences are selected and the state of the summary gets updated by $\mathrm{SentStates}$. For the word-level, all compressed word representations in a sentence are pushed to the word-level layer. In the compressive decoder, words that get selected are pushed onto the $\mathrm{WordStates}$, and once the decoder has reached the end of the sentence, it pushes the output representation of the last state onto the sentence-level layer for the next sentence. %Then the process repeats. 

%\vspace{0.1cm}

\paragraph{Extractive Decoder}
The extractive decoder selects the sentences that should go to the summary. For each sentence $s_i$ at time step $i$, the decoder takes a decision based on the encoder representation $\bs d^s_i$ and the state of the summary $\bs o^s_i$, computed as follows:
\begin{equation}
\begin{aligned}
\bs o^s_{i}=\mathrm{SentStates}(\{\bs e(c_k)\}_{k < i, z_k = 1}).
\end{aligned}
\end{equation}
\noindent where the $\bs o^s_i$ is modeled by an LSTM taking as input the already selected and compressed sentences comprising the summary so far $\{\bs e(c_k)\}_{k < i, z_k = 1}$.
This way, at each point in time, we have a representation of the summary given by the $\mathrm{SentStates}$ LSTM that encodes the state of summary generated so far, based on the past sentences already processed by the compressive decoder $\bs e(c_{i-1})$ (in $\mathrm{WordStates}$).\footnote{When using only the extractive model the summary state $\bs o_i^s$ is generated from an LSTM whose inputs correspond to the sentence encoded embeddings $\{\bs e(s_k)\}_{k < i, z_k = 1}$ instead of the previously generated compressed representations $\{\bs e(c_k)\}_{k < i, z_k = 1}$.}
\noindent The summary representation at step $i$ ($\bs o^s_{i}$) is then used to determine whether to keep or not the current sentence in the summary ($z_i=1$ or $0$ respectively). The summarizer state subsumes information about the document, sentence and summary as:
\begin{align*}
\bs {p}_{i} = \tanh (W_E[\bs d^s_{i}; \bs o^s_{i}] + \bs b^s),
\end{align*}
\noindent where $W_E$ is a model parameter, $\bs o^s_{i}$ is the dynamic LSTM state, and $\bs b^s$ is a bias term.
% hyperbolic tangent unit (tanh) nonlinearity.%\shaycomment{can you please give me the dimensions of each of the elements in the above equation?}
%\pedro{$W_e$=(512x512), \bs e(D)=512, $s_{i-1}$=512 $\bs e(s_i)$;=12 $o^s_{i\text{-}1}$=512 $\bs b^s$=512. We referred in section 7.2 these values. The representation layer operates in 512. The only transformation is the sentence encoder from 200*2 (fwd,bwd) from pre-trained embeddings to 512 and the final action to 2}

This modeling decision is crucial in order to generate variable length summaries. It captures information about the sentences or words already present in the summary, helping in better understanding the ``true'' length of the summary given the document.

Finally, the summarizer state $\bs {p}_{i}$ is used to compute the probability of the action at time $i$ as:
\begin{align*}
p(z_i \mid \bs {p}_{i}) = \frac{\exp \left( W_{z_i} \bs {p}_{i} + \bs x_{z_i} \right)}{\sum_{z' \in \{0,1\}} \exp \left( W_{z'} \bs {p}_{i} + \bs x_{z'} \right)},
\end{align*}
\noindent where $W_z$ is a model parameter and $\bs x_z$ is a bias term for the summarizer action $z$. 

We minimize the negative log-likelihood of the observed labels at training time~\cite{Dimitroff2013Weighted}, where $\lambda^s_0$ and $\lambda^s_1$ represent the distribution of each class for the given sentences:\footnote{If $M - \sum_{i=1}^M z_i {=}0$ or $\sum_{i=1}^M z_i {=}0$, we simply consider the whole term to be zero. Here $M$ represents the number of sentences in the document.}

\begin{align}
 L(\theta^s)= -\sum_{c\in \{0,1\}}\displaystyle\frac{\lambda^s_c}{\sum\limits_{i=1}^M \mathds{1}_{z_i=c}} \sum_{i, z_i = 0} \log p(z_i | \bs {p}_{i}),\displaystyle
\label{eq:loss_s}
\end{align}
\noindent where $\mathds{1}_{z_i=c}$ is the indicator function of class $c$ and $\theta^s$ represents all the training parameters of the sentence encode/decoder. At test time, the model emits probability $p(z_i\mid \bs {p}_{i})$, which is used as the soft prediction sequentially extracting the sentence $i$. We admit sentences when $p(z_i=1\mid \bs {p}_{i})>0.5$.

\vspace{0.2cm}
\paragraph{Compressive Decoder}
Our compressive decoder shares its architecture with the extractive decoder. The compressive layer is triggered every time a sentence is selected in the summary and is responsible for selecting the words within each selected sentence. In practice, $\mathrm{WordStates}$ LSTM (see Figure~\ref{fig:encdec}) is applied hierarchically after the sentence-level decoder, using as input the collected word embeddings so far:
\begin{equation}
\begin{aligned}
\bs o^w_{ij}=\mathrm{WordStates}(\{\bs e(w_{ik})\}_{k \le j, y_{ik} = 1}). \\
\end{aligned}
\end{equation}
After making the selection decision for all words pertaining to a sentence, the final state of the $\mathrm{WordStates}$, $\bs e(c_i)=\bs o^w_{iN}$
 is fed back to $\mathrm{SentStates}$ of the extractive level decoder for the consecutive sentence, as depicted in Figure \ref{fig:encdec}.
 
The word-level summarizer state representation 
depends on the encoding of words, document and sentence $\bs d^w_{ij}$, on the dynamic LSTM encoding for the summary based on the selected words ($\mathrm{WordStates}$) $\bs o^w_{i{j}}$ and sentences ($\mathrm{SentStates}$) $\bs o^s_i$:
\begin{align}
\bs {q}_{ij} = \tanh(W_C&[\bs d^w_{ij}; \bs o^s_{i}; \bs o^w_{i{j}}] + \bs b^w),
\end{align}
\noindent where $W_C$ is a model parameter and $\bs b^w$ is a bias term.
Each action at time step $j$ is computed by 
\begin{align*}
p(y_{ij} \mid \bs {q}_{ij}) = \frac{\exp \left( W_{y_{ij}} \bs {q}_{ij} + \bs x_{y_{ij}} \right)}{\sum_{y' \in \{0,1\}} \exp \left( W_{y'} \bs {q}_{ij} + \bs x_{y'} \right)},
\end{align*}
with parameter $W_{y_{ij}}$ and bias $\bs x_{y_{ij}}$.
The final loss for the compressive layer is
\begin{eqnarray}
  L(\theta^w) = \sum_{i=1}^M z_i\phi(i\mid \theta^w),
\label{eq:loss_w}
\end{eqnarray}
\noindent where $\theta^w$ represents the set of all the training parameters of the word-level encoder/decoder, $\phi(i)$ is the compressive layer loss over N words:
\begin{align}
\phi(i\mid \theta^w) = -\sum_{c\in \{0,1\}}\displaystyle\frac{\lambda^w_c}{\sum\limits_{i=1}^M \mathds{1}_{y_{ij}=c}} \sum_{i, z_i = 0} \log p(y_{ij} | \bs {q}_{ij}).
\end{align}
\noindent The total final loss is then given by the sum of the extractive and compressive counterparts, $L(\theta) = L(\theta^s) + L(\theta^w)$.
%\noindent where $N$ represents the number of words, and $\lambda^w_0,\lambda^w_1$ represent the distribution of each class for the given words.
%We minimize the negative log-likelihood of the observed labels at training time by a weighted average where the parameter $\lambda^w_0,\lambda^w_1$ represents the distribution of each class for the given words.

\section{Experimental Setup}
\label{setup}

% Following previous work in summarization \cite{Nallapati2017SummaRuNNer,See2017Get,Paulus2018Deep,Narayan2018Ranking}, 
We mainly used the CNN/DailyMail corpus \cite{Hermann2015Teaching} to evaluate our models. We used the standard splits of \newcite{Hermann2015Teaching} for training, validation, and testing (90,266/1,220/1,093 documents for CNN and 196,961/12,148/10,397 for DailyMail). To evaluate the flexibility of our model, we also evaluated our models on the Newsroom dataset~\cite{newsroom_N181065}, which includes articles form a diverse collection of sources (38 publishers) with different summary style subsets: extractive (Ext.), mixed (Mixed) and abstractive (Abs.). We used the standard splits of \newcite{newsroom_N181065} for training, validation, and testing (331,778/36,332/36,122 documents for Ext., 328,634/35,879/36,006 for Mixed and 332,554/36,380/36,522 for Abs.). We did not anonymize entities or lower case tokens.

\subsection{Estimating Oracles}
\label{susbsec:oracles}
Datasets for training extractive summarization systems do not naturally contain sentence/word-level labels. Instead, they are typically accompanied by abstractive summaries from which extraction labels are extrapolated. 
We create extractive and compressive summaries prior to training using two types of {\em oracles}.

We used an \emph{extractive oracle} to identify the set of sentences which collectively gives the highest ROUGE \cite{Lin2003Automatic} with respect to the gold summary \cite{Narayan2018Ranking}.

To build a \emph{compressive oracle}, we trained a supervised sentence labeling classifier, adapted from the Transition-Based Chunking Model \cite{Lample2016Neural}, to annotate spans in every sentence that can be dropped in the final summary. 
We used the publicly released set of 10,000 sentence-compression pairs from the Google sentence compression dataset \cite{Filippova2013Overcoming,Filippova2015Sentence} for training. After tagging all sentences in the CNN and DailyMail corpora using this compression model, we generated oracle compressive summaries based on the best average of $\mbox{ROUGE-1}$ (R1) and $\mbox{ROUGE-2}$ (R2) F$_1$ scores from the combination of all possible sentences and all removals of the marked compression chunks. 

To verify the adequacy of our proposed oracles, we show in Table~\ref{tab:score_oracle} a comparison of their scores. Our compressive oracle achieves much better scores than the extractive oracle, because of its capability to make summaries concise. Moreover, the linguistic quality of these oracles was preserved due to the tagging of the entire span by the sentence compressor trained on the sentence compression dataset.\footnote{We show examples of both oracles in Appendix~\S\ref{sec:appendix1}.} We believe that our dataset with oracle compression labels will be of significant interest to the sentence compression and summarization community. 

\begin{table}[t!]
  % \resizebox{0.8\columnwidth}{!}
  \centering
  {\footnotesize
  \begin{tabular}{| l | c c r |}
  \hline
  Oracle & R1 & R2 & RL \\ \hline
  Extractive Oracle & 54.67 & 30.37 & 50.81 \\ 
  Compressive Oracle & \textbf{57.12} & \textbf{32.59} & \textbf{53.27} \\ \hline
  \end{tabular}%
  }
  \caption{ Oracle scores obtained for the CNN and DailyMail testsets. We report ROUGE-1 (R1), ROUGE-2 (R2) and ROUGE-L (RL) F1 scores.}\label{tab:score_oracle}
\end{table}
% \footnote{Our dataset annotated with oracle compression labels is available at \url{anonymized}.}
% Additionally, we build a \emph{bag-of-words oracle} (BOW oracle), that serves as a baseline for creating a compressive oracle, in which labels are generated by simply dropping words if they do not appear in the gold summary. Unsurprisingly, oracle sentences compressed with this method are often ungrammatical.

\begin{table*}[t!]
  \setlength\tabcolsep{3.5pt}
  \center{\small
  \begin{tabular}{|l|ccc|ccc|ccc|ccc|ccc|}
    \hline 
    \multirow{2}{*}{Models}&    
    \multicolumn{3}{c|}{CNN} &
    \multicolumn{3}{c|}{DailyMail} &
    \multicolumn{3}{c|}{Newsroom Ext.} &
    \multicolumn{3}{c|}{Newsroom Mixed} &
    \multicolumn{3}{c|}{Newsroom Abs.}\\ % \cline{2-10}
     & R1 & R2 & RL & R1 & R2 & RL & R1 & R2 & RL & R1 & R2 & RL & R1 & R2 & RL    \\ \hline \hline
    \textsc{LEAD} & 29.1 & 11.1 & 25.9  & 40.7 & 18.3 & 37.2 & 53.1 & 49.0 & 52.4 & --- & --- & --- & 13.7 & 2.4 & 11.2 \\%%10.3 6.1 %%0.44 .51  \\
       %\textsc{LEAD} & 29.1 & 11.1 & 25.9 & .49 & 40.7 & 18.3 & 37.2 & .54 & 53.1 & 49.0 & 52.4 & --- & --- & --- & 13.7 & 2.4 & 11.2 \\%%10.3 6.1 %%0.44 .51  \\
   % \textsc{SummaRuNNer}~\cite{Nallapati2017SummaRuNNer}$^{\ast}$ & --- & --- & --- & --- & --- & --- & --- & ---  \\ 
    \refresh & 30.0 & 11.7  & 26.9  & 41.0  & 18.8  & 37.7  & --- & --- & --- & --- & --- & ---  & --- & --- & ---\\ %%9.7 5.5%%.44 .48 \\
    %\refresh & 30.0 & 11.7  & 26.9  & .47  & 41.0  & 18.8  & 37.7  & .49 & --- & --- & --- & --- & --- & ---  & --- & --- & ---\\ %%9.7 5.5%%.44 .48 \\
    \ExConSumm~Extractive & \textbf{32.5} & 12.6 & 28.5 & \textbf{42.8} & \textbf{19.3} & \textbf{38.9} & \textbf{69.4} & \textbf{64.3} & \textbf{68.3} & \textbf{31.9} & \textbf{16.3} & 26.9 & \textbf{17.2} & \textbf{3.1} & 13.6\\%% 12.5 7.0 %%.58.62  \\ 
    %\ExConSumm~Ext. & 32.5 & 12.6 & 28.5 &.63 & \textbf{42.8} & \textbf{19.3} & \textbf{38.9} & .67 & \textbf{69.4} & \textbf{64.3} & \textbf{68.3} & 31.9 & \textbf{16.3} & 26.9 & 17.2 & \textbf{3.1} & 13.6\\%% 12.5 7.0 %%.58.62  \\ 
    \hline \hline
    %\ExConSumm~BoW & \textbf{33.5} & 12.4 & \textbf{30.0} & 42.5 & 17.8 & 38.8 & 68.7 & 59.8 & 67.6 & \textbf{32.0} & 15.2 & \textbf{27.3} & \textbf{19.1} & 2.8 & \textbf{16.5}\\%%18.3 9.3 %%.84 .79 \\ 
    %\ExConSumm~BoW$^{\dagger}$ & \textbf{33.5} & 12.4 & \textbf{30.0} & .91 & 42.5 & 17.8 & 38.8 & .89& 68.7 & 59.8 & 67.6 & \textbf{32.0} & 15.2 & \textbf{27.3} & \textbf{19.1} & 2.8 & \textbf{16.5}\\%%18.3 9.3 %%.84 .79 \\ 
    \ExConSumm~Compressive & \textbf{32.5} & \textbf{12.7} & \textbf{29.2} & 41.7 & 18.5 & 38.4 & 68.4 & 62.9 & 67.3 & 31.7 & 16.1 & \textbf{27.0} & 17.1 & \textbf{3.1} & \textbf{14.1} \\%%13.3 6.0 %%.60 .52 \hline    
    %\ExConSumm~Comp. & 32.5 & \textbf{12.7} & 29.2 &  .67 & 41.7 & 18.5 & 38.4 & .57 & 68.4 & 62.9 & 67.3 & 31.7 & 16.1 & 27.0 & 17.1 & \textbf{3.1} & 14.1 \\%%13.3 6.0 %%.60 .52 
    
    \hline \hline 
    
    Pointer+Coverage~$^{\diamond}$& --- & --- & --- & --- & --- & --- & 39.1 & 28.0 & 36.2 & 25.5 & 11.0 & 21.1 & 14.7 & 2.3 & 11.4 \\
    
    \newcite{Tan2017Abstractive}$^{\ast}$ & 30.3 & 9.8 & 20.0  & --- & --- & --- & ---& --- & --- & ---& --- & ---& --- & --- & --- \\  
    
     \hline
  \end{tabular}
  }
  \caption{Results on the CNN, DailyMail and Newsroom test sets. We report ROUGE R1, R2 and RL F$_1$ scores. Extractive systems are in the first block, compressive in the second and abstractive in the third. We use --- whenever results are not available.  Models marked with $^{\ast}$ are not directly comparable to ours as they are based on an anonymized version of the dataset. The model marked with $^{\diamond}$ show here the results for the best configuration of \newcite{See2017Get}, referred to as Pointer{-}N in~\newcite{newsroom_N181065}, which is trained on the whole Newsroom dataset.} 
  \label{tab:corpus_results1}
\end{table*}

\subsection{Training Parameters}

% We used the pre-trained word embeddings of size 200 provided by \cite{Narayan2018Ranking}. These embeddings were trained on the One Billion Word Benchmark corpus \cite{Chelba2013One}. 

The parameters for the loss at the sentence-level were $\lambda^s_0{=}2$ and $\lambda^s_1{=}1$ and at the word-level, $\lambda^w_0{=}1$ and $\lambda^w_1{=}0.5$. We used LSTMs with $d=512$ for all hidden layers. We performed mini-batch negative log-likelihood training %(\ref{eq:loss_s}, \ref{eq:loss_w}) 
with a batch size of 2~documents for 5~training epochs.We observed the convergence of the model between the 2nd and the 3rd epochs. It took around 12 hrs on a single GTX 1080 GPU to train. We evaluated our model on the validation set after every 5,000 batches. We trained with Adam \cite{Kingma2015Adam} with an initial learning rate of~$0.001$. Our system was implemented using DyNet \cite{Neubig2017DyNet}.

\begin{table}[t!]

  \center{\footnotesize
  \begin{tabular}{|l|ccc|}
    \hline 
    \multirow{2}{*}{Models}&    
    
    \multicolumn{3}{c|}{CNN$+$DailyMail}\\ % \cline{2-10}
    & R1 & R2 & RL \\ \hline \hline %& $c_r$ 
	
    \textsc{LEAD} & 39.6 & 17.7 & 36.2 \\%6.5\\%.50 \\& .55
    \textsc{SummaRuNNer}$^{\ast}$ & 39.6 & 16.2 & 35.3 \\ %~\cite{Nallapati2017SummaRuNNer}
    \refresh & 40.0  & 18.2  & 36.6\\%5.9\\%.48 & .50
    \textsc{Latent} & 41.1 & \textbf{18.8} & 37.4 \\%{~}\cite{zhang2018neural}\\~\cite{Narayan2018Ranking}
    \ExConSumm~Extractive & \textbf{41.7} & 18.6 & {37.8} \\%7.7 \\%.63 \\ & .71
    \hline \hline
    %\ExConSumm~BoW & 41.3 & 17.3 & 37.7 \\%9.6 \\%.77\\ & .89
    \textsc{Latent+Compress} &36.7& 15.4 &34.3\\
    \ExConSumm~Compressive & 40.9 & 18.0 & 37.4\\ % 7.3 %.58 \\ \hline     & .66
    \hline \hline 
    
    Pointer+Coverage & 39.5 & 17.3 & 36.4  \\ %.87 %~& .92
    ML + RL$^{\ast}$ & 39.9 & 15.8 & 36.9 \\ %\cite{Paulus2018Deep}
    \newcite{Tan2017Abstractive}$^{\ast}$ & 38.1 & 13.9 & 34.0 \\  
    \newcite{li2018guiding} & 39.0 & 17.1 & 35.7 \\ %Key inf. guide network{~}\cite{li2018guiding}
    
    \newcite{bansal2018fastabstractive} & 40.4 & 18.0 & 37.1\\ %Saliency+Entailment{~}
    \newcite{Hsu2018Unified} & 40.7 & 18.0 & 37.1 \\%Inconsistency~loss{~}\cite{Hsu2018Unified}
    
     \newcite{Pasunuru-multireward18}& 40.9 & 17.8 & \textbf{38.5} \\%Sentence~Rewriting {~}
    \newcite{gehrmann2018bottom} & 41.2 & 18.7 & 38.3 \\%Bottom-Up{~}\cite{gehrmann2018bottom}
    % there's also Sentence Rewriting: https://arxiv.org/pdf/1805.11080.pdf
     \hline
  \end{tabular}
  }
  \caption{Results for combined CNN/DailyMail test set. }
  \label{tab:corpus_results2}
  \vspace{-0.1in}
\end{table}

\subsection{Model Evaluation}
We evaluated summarization quality using F$_1$ $\mbox{ROUGE}$ \cite{Lin2003Automatic}. We report results in terms of unigram and bigram overlap (R1) and (R2) as a means of assessing informativeness, and the longest common subsequence (RL) as a means of assessing fluency.%We report F$_1$-score, given the variable length of the summaries
\footnote{We used \texttt{pyrouge} to compute the ROUGE scores. The parameters we used were ``-a -c 95 -m -n 4 -w 1.2.''} In addition to ROUGE, which can be misleading when used as the only means to assess summaries \cite{schluter:2017:EACLshort}, we also conducted a question-answering based human evaluation to assess the informativeness of our summaries in their ability to preserve key information from the document~\cite{Narayan2018Ranking}.\footnote{We used the CNN/DailyMail QA test set of \newcite{Narayan2018Ranking} for evaluation. It includes 20 documents with a total of 71 manually written question-answer pairs.} First, questions are written using the gold summary, we then examined how many questions participants were able to answer by reading system summaries alone, without access to the article.\footnote{See Appendix~\S\ref{sec:humanQA} for more details.} Figure~\ref{fig:summaries-with-qa} shows a set of candidate summaries along with questions used for this evaluation.  

% We uploaded data in batches (one system at a time) on Mechanical Turk to ensure that the same participant does not evaluate summaries from different systems on the same set of questions.

%%%%%%%%%%%%%%%%%%%%%% ON QA dataset %%%%%%%%%%%%%%%%%%%%%55
\begin{table*}[t!]
\centering
%\resizebox{\columnwidth}{!}{%
  \center{\footnotesize
\begin{tabular}{| l | c c |c c c | c c | c c r |c|}

\hline
\multirow{3}{*}{Models} & \multicolumn{5}{c}{Bounded} & \multicolumn{6}{|c|}{Unbounded}  \\ 

 & \multicolumn{2}{c|}{Human QA} & \multicolumn{3}{c}{ROUGE}  & \multicolumn{2}{|c|}{Human QA}   & \multicolumn{3}{c|}{ROUGE} & Pearson\\ 

 & \multicolumn{1}{l}{score} & rank & R1 & R2 & RL & score & rank & R1 & R2 & RL & r \\ \hline
	
\multicolumn{1}{|l|}{\textsc{LEAD}} & 
	% BOUNDED
	25.50 & 4\textsuperscript{rd} & 
	30.9 & 11.9 & 29.1 &%43.9&
	% UNBOUNDED
	\multicolumn{1}{l}{36.33} & 5\textsuperscript{th} &  
	31.6 & 13.5 & 29.3 & 0.40\\ % & 0.66 \\ %76.3
	
\multicolumn{1}{|l|}{\refresh\ }  & 
	% BOUNDED
	 \multicolumn{1}{c}{20.88} & 6\textsuperscript{th}  & 
	 37.4 & 17.3 & 34.8 &%45.2&
	% UNBOUNDED
	\textbf{66.34} & 1\textsuperscript{st}    & 
	\textbf{43.8} & \textbf{25.8} & \textbf{41.6} & 0.60 \\ % & 0.56 \\ %76.0

\multicolumn{1}{|l|}{\textsc{Latent}} & 
	% BOUNDED
	38.45 & 2\textsuperscript{nd} & 
	\textbf{38.9} & 19.6 & 36.4  &%43.3&
	% UNBOUNDED
	\multicolumn{1}{l}{53.38} & 4\textsuperscript{th} &  
	40.7 & 22.0 & 38.1 & -0.02\\ % & 0.60\\ 

\multicolumn{1}{|l|}{\ExConSumm\ Extractive }  & 
	% BOUNDED
	\multicolumn{1}{c}{36.34} & 3\textsuperscript{rd}  & 
	38.4 & 18.5 & 35.9 &%43.9 &
	% UNBOUNDED
	54.93 & 3\textsuperscript{rd} & 
	40.8 & 21.0 & 38.2 & 0.68 \\ % & 0.68 \\ %56.2

\multicolumn{1}{|l|}{\ExConSumm\ Compressive} & 
	% BOUNDED
	 \textbf{39.44} & \textsc{1\textsuperscript{st}} & 
	38.8 & 19.0 & \textbf{37.0} &%45.3 &
	% UNBOUNDED
	\multicolumn{1}{l}{57.32} & 2\textsuperscript{nd} &
	41.4 & 22.6 & 39.1 & \textbf{0.72}\\ %  & 0.65 \\

%\multicolumn{1}{|l|}{\ExConSumm\ BoW} & 

	% BOUNDED
%	31.69 &  4\textsuperscript{th} & \textbf{39.2} & 18.2 & \textbf{37.0}  & \textbf{1.16}&%38.9&		\multicolumn{1}{c|}{} &	
	
	% UNDBOUNDED
	%\multicolumn{1}{l|}{36.06} & 6\textsuperscript{th} & 41.3 & 20.1 & 38.7  & \textbf{0.97}\\\cline{1-7} \cline{9-14}%38.9 \cline{1-7} \cline{9-14}

\multicolumn{1}{|l|}{Pointer+Coverage} & 
	%~\cite{See2017Get}
	% BOUNDED
	\multicolumn{1}{c}{24.51} & 5\textsuperscript{th}  & 
	38.4 & \textbf{19.7} & 36.7 &%40.4&
	% UNDBOUNDED
	28.73 & 6\textsuperscript{th}   & 
	40.2 & 21.4 & 38.0 & 0.30\\ \hline % & 0.94  \\\cline{1-6} \cline{8-13}

\end{tabular}
%}%resize box
}
  \caption{ QA evaluations: limited length (Bounded) and full length (Unbounded) summaries. We also show ROUGE scores for the summaries being evaluated. We report the Pearson correlation coefficient between the human and predicted summary lengths\label{tab:heval}}
  % The gold summaries have $\bar{L}$=$45.5$ avg. length. %$\textrm{CR}$ shows the compression ratio (higher produces shorter summaries) \label{tab:heval}}
  \vspace{-0.1in}
\end{table*}

\subsection{Model and Baselines} 
We evaluated our model \ExConSumm\ in two settings: Extractive (selects sentences to assemble the summary) and Compressive (selects sentences and compresses them by removing unnecessary spans of words). 
We compared our models against a baseline (\textsc{LEAD}) that selects the first $m$~leading sentences from each document,\footnote{We follow \newcite{Narayan2018Ranking} and set $m=3$ for CNN and $4$ for DailyMail.  We follow \newcite{newsroom_N181065} and set $m=2$ for Newsroom.} three neural extractive models, and various abstractive models. For the extractive models, we used \textsc{SummaRuNNer}~\cite{Nallapati2017SummaRuNNer}, since it shares some similarity to our model, %and the state-of-the-art extractive models
\refresh~\cite{Narayan2018Ranking} trained with reinforcement learning and \textsc{Latent}~\cite{zhang2018neural} a neural architecture that makes use of latent variable to avoid creating oracle summaries. We further compare against \textsc{Latent+Compress}~\cite{zhang2018neural}, an extension of the \textsc{Latent} model that learns to map extracted sentences
to final summaries using an attention-based seq2seq model 
\cite{Rush2015Neural}.
All models, unlike ours, extract a fixed number of sentences to assemble their summaries. 
For abstractive models, we compare against the state-of-the art models of \textsc{Pointer+Coverage}~\cite{See2017Get}, \textsc{ML+RL}~\cite{Paulus2018Deep}, and \newcite{Tan2017Abstractive} among others.

\section{Results}
\label{sec:results}
\subsection{Automatic Evaluation}
Table~\ref{tab:corpus_results1} and \ref{tab:corpus_results2} show results for the evaluations on the CNN/DailyMail and Newsroom test sets.

\paragraph{Comparison with Extractive Systems.} 
\ExConSumm{~}Compressive performs best on the CNN dataset and \ExConSumm{~}Extractive on the DailyMail dataset, probably due to the fact that the CNN dataset is less biased towards extractive methods than the DailyMail dataset \cite{narayan18:xsum}. We report similar results on the Newsroom dataset. \ExConSumm{~}Compressive tends to perform better for mixed (Mixed) and abstractive (Abs.) subsets, while \ExConSumm{~}Extractive performs better for the extractive (Ext.) subset. Our experiments demonstrate that our compressive model tends to perform better on the dataset which promotes abstractive summaries.

We find that \ExConSumm~Extractive consistently performs better on all metrics when compared to any of the other extractive models, except for the single case where it is narrowly behind \textsc{Latent} on R2 (18.6 vs 18.8) for the CNN/DailyMail combined test set. It even outperforms \refresh, which is trained with reinforcement learning. We hypothesize that its superior performance stems from the ability to generate variable length summaries. \refresh\ or \textsc{Latent}, on the other hand, always produces a fixed length summary. 

\paragraph{Comparison with Compressive System.}
\ExConSumm{~}Compressive reports superior performance compared to \textsc{Latent+Compress} (+4.2 for R1, +2.6 for R2 and +3.1 for RL). Our results demonstrate that our compressive system is more suitable for document summarization. It first selects sentences and then compresses them by removing irrelevant spans of words. It makes use of an advance oracle sentence compressor trained on a dedicated sentence compression dataset (Sec.~\ref{susbsec:oracles}).  
In contrast, \textsc{Latent+Compress} naively trains a sequence-to-sequence compressor to map a sentence in the document to a sentence in the summary. 

\begin{figure}[t!]
    \begin{subfigure}[b]{\columnwidth}
        \includegraphics[trim={0.15in, 0in, 0in, 0in},scale=0.33,clip]{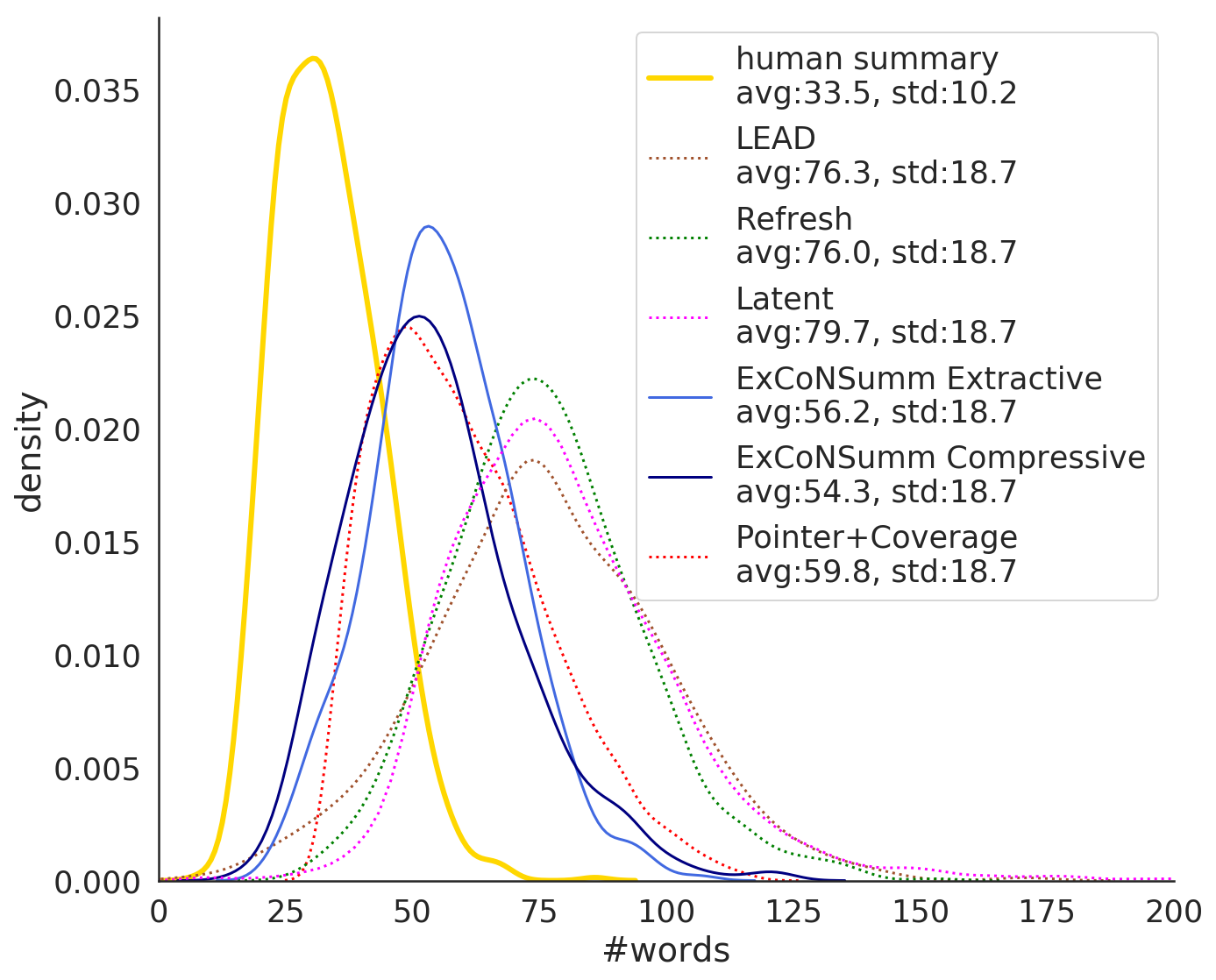}
    \end{subfigure}
    \caption{Word distribution in comparison with the human summaries for CNN dataset. Density curves show the length distributions of human authored and system produced summaries. }
    \label{fig:word_distribution}
      \vspace{-0.2in}
\end{figure}

\paragraph{Comparison with Abstractive Systems.}
Both \ExConSumm{~}Extractive and{~}Compressive outperform most of the abstractive systems including Pointer+Coverage \cite{See2017Get}. When comparing with more recent methods~\cite{Pasunuru-multireward18,gehrmann2018bottom}, our model has comparable performance.

% On the CNN dataset, whose gold summaries are more compressive in nature, we obtain the best results with our our compressive model, whereas on the DailyMail, whose gold summaries are more extractive, our extractive performs best.
% The Newsroom dataset contains three partitions: an extractive, an abstractive and a mixed partition. One of our models performs best in all three partitions. \ExConSumm{~}Compressive tends to perform better for the mixed and abstractive datasets, while \ExConSumm{~}Extractive performs better for the extractive dataset.
% % In the rest of the paper, we refer to our model as \ExConSumm.
% % We show results for both the extractive and compressive variants of \ExConSumm. 

% \paragraph{Comparison with Extractive Systems}

% We find that \ExConSumm~Extractive performs better on all metrics when compared to any of the other extractive models, even outperforming \refresh, which is trained with reinforcement learning. We hypothesize that its superior performance stems from the ability to generate variable length summaries. \refresh, on the other hand, always produces a fixed length summary. 

% \paragraph{Comparison with Abstractive Systems}
%  Our compressive model has the best score when trained separately on CNN and even on the abstractive dataset of Newsroom (Table~\ref{tab:corpus_results1}). For the combined dataset (CNN+DailyMail), %, our compressive model perfo
rmed almost as well as the best abstractive model.
% compared with other abstractive systems such as~\newcite{See2017Get} (\textsc{Pointer+Coverage}), we obtain better results, but 
% when comparing with more recent methods~\cite{Pasunuru-multireward18,zhang2018neural,gehrmann2018bottom} our model has comparable performance. 

\begin{figure}[t!]
  \center{\fontsize{8.5}{7}\selectfont % \scriptsize \footnotesize
    \begin{tabular}{p{7.5cm}}
      \hline \vspace{0.1pt}
        \textbf{\textsc{LEAD}} \\
        \textbullet \hspace{0.1cm} (CNN) A top al Qaeda in the Arabian Peninsula leader--who a few years ago was in a U.S. detention facility--was among five killed in an airstrike in Yemen, the terror group said, showing the organization is vulnerable even as Yemen appears close to civil war.  \\ 
        \textbullet \hspace{0.1cm} Ibrahim al-Rubaish died Monday night in what \textcolor{forestgreen}{AQAP}'s media wing, Al-Malahem Media, called a ``\textcolor{sepia}{crusader airstrike}.''  \\
        \textbullet \hspace{0.1cm} The Al-Malahem Media obituary characterized al-Rubaish as a religious scholar and combat commander. \\ \vspace{0.1pt}
        \textbf{\textsc{Refresh}} \\
        \textbullet \hspace{0.1cm} (CNN) A top al Qaeda in the Arabian Peninsula leader--who a few years ago was in a U.S. detention facility--was among five killed in an airstrike in Yemen, the terror group said, showing the organization is vulnerable even as Yemen appears close to civil war. \\
        \textbullet \hspace{0.1cm} Ibrahim al-Rubaish died Monday night in what \textcolor{forestgreen}{AQAP}'s media wing, Al-Malahem Media, called a ``\textcolor{sepia}{crusader airstrike}.'' \\
        \textbullet \hspace{0.1cm} Al-Rubaish was once held by the U.S. government at its detention facility in \textcolor{orangered}{Guantanamo Bay}, Cuba. \\ \vspace{0.1pt}
        \textbf{\textsc{Latent}} \\
        \textbullet \hspace{0.1cm} (CNN) A top al Qaeda in the Arabian Peninsula leader--who a few years ago was in a U.S. detention facility--was among five killed in an airstrike in Yemen, the terror group said, showing the organization is vulnerable even as Yemen appears close to civil war. \\
        \textbullet \hspace{0.1cm} Ibrahim al-Rubaish died Monday night in what \textcolor{forestgreen}{AQAP}'s media wing, Al-Malahem Media, called a ``\textcolor{sepia}{crusader airstrike}.'' The Al-Malahem Media obituary characterized al-Rubaish as a religious scholar and combat commander. \\
        \textbullet \hspace{0.1cm} A Yemeni Defense Ministry official and two Yemeni national security officials not authorized to speak on record confirmed that al-Rubaish had been killed, but could not specify how he died. \\ \vspace{0.1pt}
        
        \textbf{\ExConSumm\ Extractive} \\
        \textbullet \hspace{0.1cm} (CNN) A top al Qaeda in the Arabian Peninsula leader--who a few years ago was in a U.S. detention facility--was among five killed in an airstrike in Yemen, the terror group said, showing the organization is vulnerable even as Yemen appears close to civil war. \\
        \textbullet \hspace{0.1cm} Ibrahim al-Rubaish died Monday night in what \textcolor{forestgreen}{AQAP}'s media wing, Al-Malahem Media, called a ``\textcolor{sepia}{crusader airstrike}.'' \\ \hline \vspace{0.1pt}
        
        \textbf{\ExConSumm\ Compressive} \\
        \textbullet \hspace{0.1cm} A top al Qaeda in the Arabian Peninsula leader--who a few years ago was in a U.S. detention facility--was among five killed in an airstrike in Yemen. 
        \textbullet \hspace{0.1cm} Ibrahim al-Rubaish died in what \textcolor{forestgreen}{AQAP}'s media wing, Al-Malahem Media, called a ``\textcolor{sepia}{crusader airstrike}.'' \\ \hline   \vspace{0.1pt}  
      
        \textbf{Pointer+Coverage} \\
        \textbullet \hspace{0.1cm} Ibrahim al-Rubaish was among a number of detainees who sued the administration of then-president George W. Bush to challenge the legality of their confinement in Gitmo. 
        \textbullet \hspace{0.1cm} al-Rubaish was once held by the U.S. government at its detention facility in \textcolor{orangered}{Guantanamo bay}, Cuba. \\ \hline \vspace{0.1pt}

        \textbf{\textsc{GOLD}} \\
        \textbullet \hspace{0.1cm} \textcolor{forestgreen}{AQAP} says a ``\textcolor{sepia}{crusader airstrike}'' killed Ibrahim al-Rubaish 
        \textbullet \hspace{0.1cm} Al-Rubaish was once detained by the United States in \textcolor{orangered}{Guantanamo} \\ \hline \vspace{0.1pt}
        
        \textbf{Question-Answer Pairs} \\
        \textbullet \hspace{0.1cm} Who said that an airstrike killed Ibrahim al-Rubaish? (\textcolor{forestgreen}{AQAP}) 
        \textbullet \hspace{0.1cm} What was the airstrike called? (\textcolor{sepia}{crusader airstrike}) 
        \textbullet \hspace{0.1cm} Where was Ibrahim al-Rubaish once detained? (\textcolor{orangered}{Guantanamo}) \\ \hline
    \end{tabular}
    }
  \caption{ Example output summaries on the CNN/DailyMail dataset, gold standard summary, and corresponding questions. The questions are manually written using the \textsc{gold} summary. The same \ExConSumm\ summaries are shown in Figure~\ref{fig:summaries-intro}, but the strike-through spans are now removed.}\label{fig:summaries-with-qa}
  \vspace{-0.5cm}
\end{figure}

\paragraph{Summary Versatility.}
We evaluate the ability of our model to generate variable length summaries. Table~\ref{tab:heval} show the Pearson correlation coefficient between the lengths of the human generated summaries against each unbounded model. Our compressive approach obtains the best results, with a Pearson correlation coefficient of 0.72 ($p<0.001$).

Figure~\ref{fig:word_distribution} also shows the distribution of words per summary for the models where predictions were available. Interestingly, both \ExConSumm{~}Extractive and{~}Compressive follow the human distribution much better than other extractive systems (\textsc{Lead}, \refresh\ and \textsc{Latent}), since they are able to generate variable-length summaries depending on the input text. Our compressive model generates a word distribution much closer to the abstractive Pointer+Coverage model but achieves better compression ratio; the summaries generated by Pointer+Coverage contain 59.8
words, while those generated by \ExConSumm{~}Compressive have 54.3 words on average.

\subsection{QA Evaluation}
Table~\ref{tab:heval} shows results from our question answering based human evaluation.  We elicited human judgements in two settings: the ``Unbounded'', where participants were shown the full system produced summaries; and  the ``Bounded'', where participants were shown summaries that were limited to the same size as the gold summaries. 

For the ``Unbounded'' setting, the output summaries produced by \refresh\, were able to answer most of the questions correctly, our Compressive and~Extractive systems were placed at the 2nd and 3rd places respectively.\footnote{We carried out pairwise comparisons
between all models to assess whether system differences are statistically significant. We found that there is no statistically significant difference between \refresh\ and \ExConSumm~Compressive. We use a one-way ANOVA with posthoc Tukey HSD tests with $p<0.01$. The differences among \textsc{Latent} and both variants of \ExConSumm, and between \textsc{lead} and Pointer+Coverage are also statistically insignificant. All
other differences are statistically significant.}

We observed that our systems were able to produce more concise summaries than those produced by \refresh\ (avg. length in words: 76.0 for \refresh, 56.2 for \ExConSumm~Extractive and 54.3 for \ExConSumm~Compressive; see Figure~\ref{fig:word_distribution}). \refresh\ is prone to generating verbose summaries, consequently it has an advantage of accumulating more information. In the ``Bounded'' setting, we aim to reduce this unfair advantage. Scores are overall lower since the summary sizes are truncated to gold size. The \ExConSumm~Compressive summaries rank first and can answer 39.44\% of questions correctly. \ExConSumm~Extractive retains its 3rd place answering 36.34\% of questions correctly.\footnote{The differences among both variants of \ExConSumm\ and \textsc{Latent}, and among \textsc{lead}, \refresh\ and Pointer+Coverage are statistically insignificant. All other differences are statistically significant. We use a one-way ANOVA with posthoc Tukey HSD tests with $p<0.01$.} These results demonstrate that our models generate concise and informative summaries that correlate well with the human summary lengths.\footnote{App.~\S\ref{sec:humanQA} shows more examples of our summaries.}

\subsection{Summary State Representation}
Next, we performed an ablation study to investigate the importance of the summary state representation $\bs o^s_i$ w.r.t. the quality of the overall summary. We tested against a \textsc{State averaging} variant, where we replace $\bs o^s_i$ by a weighted average, analogous to~\newcite{Nallapati2017SummaRuNNer}, $\bs o^{avg\ s}_i= \sum_{i=1}^{j-1}\bs e(s_i)p(z_i\mid \bs {p}_{i}^{avg})$, where $\bs {p}_{i}^{avg}$ has the same form as $\bs {p}_{i}$ but depends recursively on the previous summary state $\bs o^{avg\ s}_{i-1}$. Table~\ref{tab:ablation} shows that using an LSTM state $\bs o^s_i$ to model the current sentences in the summary is very important.
The other ablation study shows how learning to extract and compress in a disjoint approach (\ExConSumm{ Ext+Comp oracle}) performs against a joint learning approach (\ExConSumm{ Compressive}). We compared summaries generated from our best extractive model and compressed them with a compressive oracle. Our joint learning model achieves the best performance in all metrics compared with the other ablations, suggesting that  joint learning and using a summary state representation is beneficial for summarization.
% compared to keeping a weighted average over all previous sentences (\textsc{RUN}) and even more than not having this dependency at all (\textsc{no state}). 

\begin{table}[t!]
  \center{\footnotesize
    \begin{tabular}{|@{~}l@{~~}| l l r | } 
      \hline
      \multirow{2}{*}{State} & \multicolumn{3}{c|}{ROUGE} \\
       & R1 & R2 & RL\\ \hline 
      \ExConSumm~Extractive & \textbf{32.5} & \textbf{12.6} & \textbf{28.5}\\
      \textsc{State averaging}        & 30.0 & 12.3 & 26.9\\
    \hline \hline
    \ExConSumm{ Compressive}         & \textbf{32.5} & \textbf{12.7} & \textbf{29.2}\\
      \ExConSumm~Ext+Comp oracle        & {25.5} & 9.3 & {23.7}\\
      \hline
    \end{tabular}}
  \caption{ Summary state ablation for the CNN dataset.\label{tab:ablation}}%See text for more detail.} 
  %\vspace{-0.2in}
\end{table}

\section{Conclusions}
\label{sec:conclusion}
We developed \ExConSumm, a novel summarization model to generate variable length extractive and compressive summaries. 
Experimental results show that the ability of our model to learn a dynamic representation of the summary produces summaries that are informative, concise, and correlate well with human generated summary lengths. 
Our model outperforms state-of-the-art extractive and most of abstractive systems on the CNN and DailyMail  datasets, when evaluated automatically, and through human evaluation for the bounded scenario. We further obtain state-of-the-art results on Newsroom, a more abstractive summary dataset.

% We propose \ExConSumm: a novel summarization architecture which operates by learning dynamic representations of the summary being generated, and is capable of both sentence extraction and compression, achieving a good balance between sequence selection and coherence. We describe a novel training method with compressive oracles, and release a new CNN/DailyMail dataset with the corresponding annotations. Our models outpreform state-of-the-art methods on the $\mbox{ROUGE-1}$, $\mbox{ROUGE-2}$ and  $\mbox{ROUGE-L}$ metrics in all standard CNN/DailyMail configurations. Furthermore, we also perform human evaluation over the generated summaries, and conclude that \sebastiao{...waiting for human results}. As our model is able to stack hierarchically, one future direction could be exploring the addition of another layer to represent the relations among documents in a multi-document scenario.

% A possible future direction would be to check how much we could further gain if we trained our model with reinforcement learning as well. \pedro{move this to conclusion/future work?}

%  A possible future use of this model is as an encoder in a seq2seq model to produce better abstractive summaries.

%\subsection{Future Work}
%\sebastiao{optional:}
%\begin{itemize}
%\item As our model is able to stack hierarchically, one future direction should explore the addition of another layer to represent the relations among document in a multi-document scenario.
%\item Explore if our model would benefit from training with reinforcement learning.
%\end{itemize}

\section*{Acknowledgments}
This work is supported by the EU H2020 SUMMA project (grant agreement N\textsuperscript{\underline{o}}~688139), by Lisbon Regional Operational Programme (Lisboa 2020), under the Portugal 2020 Partnership Agreement, through the European Regional Development Fund (ERDF), within project INSIGHT (N\textsuperscript{\underline{o}}~033869), 
by the European Research Council (ERC StG DeepSPIN 758969),
and by the Funda\c{c}\~ao para a Ci\^encia e Tecnologia through contracts UID/EEA/50008/2019 and CMUPERI/TIC/0046/2014 (GoLocal).

\bibliography{naaclhlt2019}
\bibliographystyle{acl_natbib_nourl}

\appendix
\cleardoublepage

\section{Appendices}
\label{sec:appendix}

\subsection{Estimating Summary Oracles}
\label{sec:appendix1}

We describe our method to estimate extractive and compressive oracle summaries prior to training using two types of {\em oracles}. We build these oracles in order to train our model with a supervised objective by minimizing a negative log-likelihood function. We create documents annotated with sentence-level and word-level extraction labels, which correspond to the gold values of both variables $z_i$ and $y_{ij}$ respectively.

\paragraph{Extractive Oracle.}
We followed \newcite{Narayan2018Ranking} and identified the set of sentences which collectively give the highest ROUGE \cite{Lin2003Automatic} with respect to the gold summary. More concretely, we assembled candidate summaries efficiently by first selecting $p$ sentences from the document which on their own have high ROUGE scores. We then generated all possible combinations of $p$ sentences subject to maximum length $m$ ($3$ for CNN and $4$ for DailyMail) and evaluated them against the gold summary. We select the summary with the highest mean of ROUGE-1, ROUGE-2, and ROUGE-L F1 scores.

% \subsection{Bag of Words Compressive Training}
% \label{sec:bowtraining}

% As a baseline for our compressive experiments, we trained a compressive model using the above oracles for sentence selection, but instead of feeding the compressive layer with spans to be removed from the sentences, we fed the compressive layer with the bag-of-words (BoW) of the human gold summary and guided the training with $z{=}1$ when the word exists in the gold summary and $z{=}0$ otherwise. 

% \afonso{Must decide if we should keep this here or move to results and discussion}
% \sebastiao{if we move, we must remove the back-reference i added in results section}
% Later, we'll evaluate this model in terms of gramaticality and coherence against the model trained with compressed sentences. The results will show our initial assumption that the summary LSTM provides gramaticality and coherence at the word level. We could not devise an experiment at the sentence layer, but we assume that this experiment at the word-level would hold at the sentence level.

\begin{figure}[t!]
  \center{\fontsize{8.5}{6.2}\selectfont % \scriptsize \footnotesize
    \begin{tabular}{p{7.5cm}}
      \hline \vspace{0.1pt}
        \textbf{\textsc{Extractive Oracle}}\\
        \textbullet \hspace{0.1cm} Wijnaldum - who has netted 16 goals in all competitions for the Dutch giants this season - has been linked with a move to Manchester United, Arsenal and Newcastle. 
        \textbullet \hspace{0.1cm} The PSV captain could help his club help end their seven-year wait for the Eredivisie title if they beat Heerenveen at home on Saturday.\\
        \vspace{0.1pt}
        
        \textbf{\textsc{Compressive Oracle}}\\
        \textbullet \hspace{0.1cm}  \ST{PSV Eindhoven midfielder } Georginio Wijnaldum \ST{ has } \ST{ fueled speculation that he } is eyeing a move to the Premier League in the summer. 
        \textbullet \hspace{0.1cm} Wijnaldum \ST{ - } who has netted 16 goals in all competitions for the Dutch giants this season - has been linked with a move to Manchester United, Arsenal and Newcastle.\\
        \vspace{0.1pt}
        
        \textbf{\textsc{BOW Oracle} (Baseline)}\\
        \textbullet \hspace{0.1cm}
        Wijnaldum \ST{- who} has \ST{netted} 16 goals in all competitions for \ST{the} Dutch giants this \ST{season -} has been linked with a \ST{move} to Manchester United , Arsenal and Newcastle \textbullet \hspace{0.1cm} \ST{The} PSV \ST{captain could help his club help end} their \ST{seven-year wait} for \ST{the} Eredivisie title \ST{if they beat} Heerenveen \ST{at home} on Saturday\\
        \vspace{0.1pt}
        
        \textbf{\textsc{GOLD}} \\
        \textbullet \hspace{0.1cm} Georginio Wijnaldum is set to guide PSV to their first title in seven years 
        \textbullet \hspace{0.1cm} Dutch giants can win Eredivisie with a win over Heerenveen on Saturday
        \textbullet \hspace{0.1cm} Manchester United, Arsenal and Newcastle have been linked with him \textbullet \hspace{0.1cm}  Midfielder has scored 16 goals in all competitions for PSV this term\\
        \hline \vspace{0.1pt}
        
        \textbf{\textsc{Extractive Oracle}}\\
        \textbullet \hspace{0.1cm}  Wrap a filter around the centre of an ice-cream cone to keep little hands clean from ice-cream mess. 
        \textbullet \hspace{0.1cm} Apply a dab of your favourite shoe polish on the filter and use it as an applicator. 
        \textbullet \hspace{0.1cm} As they are lint-free coffee filters can be used to polish glass, and clean mirrors. \\
        \vspace{0.1pt}
        
        \textbf{\textsc{Compressive Oracle}}\\
        \textbullet \hspace{0.1cm} Keep celery crispy \ST{ by storing them with a coffee filter, } \ST{ which are far more absorbent than kitchen roll, and will } \ST{ absorb moisture from the vegetables }. 
        \textbullet \hspace{0.1cm} Wrap a filter around the centre of an ice-cream cone to keep \ST{ little } hands clean from ice-cream mess. 
        \textbullet \hspace{0.1cm} Use it as a bouquet garni holder \ST{ the next time you're } \ST{ making soup or stews }. 
        \textbullet \hspace{0.1cm} Keep clothes smelling fresh \\
        \textbullet \hspace{0.1cm} Coffee filters are perfect for polishing leather shoes \ST{ as they are lint-free and so won't leave unsightly streak } \ST{ marks on your shoes }.\\
        \vspace{0.1pt}
        
        \textbf{\textsc{BOW Oracle} (Baseline)}\\
        \textbullet \hspace{0.1cm}
        \ST{Stop messy ice-cream }\ST{spillage} Wrap a \ST{filter} around \ST{the centre} of \ST{an ice-cream cone} to keep \ST{little} \ST{hands }\ST{clean }\ST{from }\ST{ice-cream }\ST{ mess} \textbullet \hspace{0.1cm} \ST{Apply }\ST{a }\ST{ dab} of \ST{your} \ST{favourite }\ST{shoe} polish \ST{on the filter} and \ST{use} it as \ST{an applicator} \textbullet \hspace{0.1cm}\ST{As they} are \ST{lint-free coffee} filters \ST{can be used} to polish \ST{glass}, and \ST{clean mirrors}\\
        \vspace{0.1pt}
        
        \textbf{\textsc{GOLD}} \\
        \textbullet \hspace{0.1cm} Lint-free and tear resistant filters are good for a range of household tasks 
        \textbullet \hspace{0.1cm} Wrap one sheet around celery stalks when storing in fridge to keep crisp
        \textbullet \hspace{0.1cm} Use it to polish shoes, keep laundry smelling fresh and even as a plate \\
        \hline
    \end{tabular}
    }
  \caption{ Examples of our estimated oracle summaries along with the reference summary for the CNN and DailyMail datasets. For illustration, the compressive oracle shows the removed spans strike-through.}\label{fig:oracles-example}
\end{figure}

\begin{table*}[t!]
  \setlength\tabcolsep{3.5pt}
  \center{\footnotesize
  \begin{tabular}{|l|ccc|ccc|ccc|ccc|ccc|}
    \hline 
    \multirow{2}{*}{Models} & \multicolumn{3}{c|}{CNN} & 
    \multicolumn{3}{c|}{DailyMail} &
    \multicolumn{3}{c|}{Newsroom Ext.} &
    \multicolumn{3}{c|}{Newsroom Mixed} &
    \multicolumn{3}{c|}{Newsroom Abs.}\\ % \cline{2-10}
     & R1 & R2 & RL & R1 & R2 & RL & R1 & R2 & RL & R1 & R2 & RL & R1 & R2 & RL    \\ \hline \hline
    \ExConSumm~Extractive & 32.5 & 12.6 & 28.5 & \textbf{42.8} & \textbf{19.3} & \textbf{38.9} & \textbf{69.4} & \textbf{64.3} & \textbf{68.3} & 31.9 & \textbf{16.3} & 26.9 & 17.2 & \textbf{3.1} & 13.6\\%% 12.5 7.0 %%.58.62  \\ 
    \ExConSumm~BOW & \textbf{33.5} & 12.4 & \textbf{30.0} & 42.5 & 17.8 & 38.8 & 68.7 & 59.8 & 67.6 & \textbf{32.0} & 15.2 & \textbf{27.3} & \textbf{19.1} & 2.8 & \textbf{16.5}\\%%18.3 9.3 %%.84.79 \\ 
     \ExConSumm~Compressive & 32.5 & \textbf{12.7} & 29.2 & 41.7 & 18.5 & 38.4 & 68.4 & 62.9 & 67.3 & 31.7 & 16.1 & 27.0 & 17.1 & \textbf{3.1} & 14.1 \\%%13.3 6.0 %%.60 .52 \hline    
     \hline
  \end{tabular}
  }
  \caption{ Comparison of BOW model against Extractive and Compressive models. Most of the numbers are repeated from Table~\ref{tab:corpus_results1}.} 
  \label{tab:corpus_resultsO}
\end{table*}

\paragraph{Compressive Oracle.}
The primary challenge in building a compressive oracle lies in preserving the grammaticality of compressed sentences. Following the sentence compression literature \cite{McDonald2006Discriminative, Clarke2008Global, Berg-Kirkpatrick2011Jointly, Filippova2013Overcoming, Filippova2015Sentence}, we train a supervised neural model to annotate spans in every sentence that can be dropped. In particular, we trained a supervised sentence labeling classifier adapted from \newcite{Lample2016Neural}. To train our classifier, we used the publicly released set of 10,000 sentence-compression pairs from the Google sentence compression dataset \cite{Filippova2015Sentence, Filippova2013Overcoming}. We removed the first 1,000 sentences as the development set and used the remaining ones as the training set. 

After training our classifier for 30 epochs, it achieved a per-sentence accuracy of 21\%, a word-based F-1 score of 78\% and a compression ratio of 0.38. The parameters for the model were: 2 layers, dropout of 0.1, hidden dimension of size 400, action dimension of 20 and relation dimension of 20. 
We used the One Billion Word Benchmark corpus \cite{Chelba2013One} to train word embeddings with the skip-gram model \cite{Mikolov2013Distributed} using context window size 6, negative sampling size 10, and hierarchical softmax 1. Same embeddings were used to train our summarization model also. 
For details of the evaluation metrics, please see~\newcite{Filippova2015Sentence}.

After tagging all sentences in the CNN and DailyMail corpora using this compression model, we generated oracle compressive summaries based on the best average of $\mbox{ROUGE-1}$ and $\mbox{ROUGE-2}$ F$_1$ scores from the combination of all possible sentences with all combinations of removals of the marked compression chunks. To solve this combinatorial problem, our algorithm recursively selects the $n$ possible sentences with the best accumulated score. Due to performance reasons, we used a simplified algorithm which uses unigrams and bigrams overlap computed incrementally at each recursion level, instead of the official ROUGE metric \cite{Lin2003Automatic}. 
% We stopped the recursion at maximum seven sentences and the best path was selected, thus allowing a maximum of seven compressed sentences per oracle.
We only allow a maximum of seven compressed sentences per oracle.
Our compressive oracle achieves much better scores than the extractive oracle because of its capability to make summaries concise (see Table~\ref{tab:score_oracle} in the paper). Moreover, the linguistic quality of these oracles was preserved due to the tagging of the entire span by the sentence compressor. See Figure~\ref{fig:oracles-example} for examples of our oracle summaries. 

\begin{figure}[t!]
  \center{\fontsize{8.5}{6.2}\selectfont
    \begin{tabular}{| p{7.2cm} |}
      \hline 
      \textbf{\ExConSumm\ Extractive}\\
      \textbullet \hspace{0.1cm} Beverly Hills Police investigated an incident in January 2014 in which West was accused of assaulting a man at a Beverly Hills chiropractor's office. \\
      \textbullet \hspace{0.1cm} The photographer, Daniel Ramos, had filed the civil suit against West after the hip-hop star attacked him and tried to wrestle his camera from him in July 2013 at Los Angeles International Airport. \\
      \vspace{0.1pt}
      \textbf{\ExConSumm\ BOW (Baseline)} \\
      \textbullet \hspace{0.1cm}   \sout{\textcolor{purple}{ (CNN) }} Kanye West has settled a lawsuit with a paparazzi photographer he assaulted \sout{\textcolor{purple}{ -- }} and the two have  \sout{\textcolor{purple}{ shaken }} on  \sout{\textcolor{purple}{ it }}. \\
      \textbullet \hspace{0.1cm} The photographer, Daniel Ramos, had filed the civil suit against West \sout{\textcolor{purple}{ after }} the \sout{\textcolor{purple}{ hip-hop star attacked him }} and \sout{\textcolor{purple}{ tried }} to \sout{\textcolor{purple}{ wrestle }} his \sout{\textcolor{purple}{ camera from him }} in \sout{\textcolor{purple}{ July 2013 }} at \sout{\textcolor{purple}{ Los }} \sout{\textcolor{purple}{ Angeles International Airport }}. \\
      \textbullet \hspace{0.1cm} West pleaded no contest last year to \sout{\textcolor{purple}{ a misdemeanor count }} \sout{\textcolor{purple}{ of }} battery \sout{\textcolor{purple}{ over the scuffle }}. \\
      
      \vspace{0.1pt}
      \textbf{\ExConSumm\ Compressive} \\
      \textbullet \hspace{0.1cm}  \sout{\textcolor{purple}{(CNN)}} Kanye West has settled a lawsuit with a paparazzi photographer he assaulted--and the two have shaken on it. \\
      \textbullet \hspace{0.1cm} The photographer, Daniel Ramos, had filed the civil suit against West \sout{\textcolor{purple}{ after the hip-hop star attacked him and tried to}} \sout{\textcolor{purple}{ wrestle his camera from him in July 2013 at Los Angeles}} \sout{\textcolor{purple}{International Airport}}. \\ 
      \textbullet \hspace{0.1cm} West pleaded no contest \sout{\textcolor{purple}{last year}} to a misdemeanor count of battery \sout{\textcolor{purple}{over the scuffle}}. \\ \hline
    \end{tabular}
    }
  \caption{ Summaries produced by the \ExConSumm\ Extractive, BoW compressive and compressive methods. For illustration, BoW and compressive summaries show the removed spans strike-through.}\label{fig:summaries-with-BOW}
    % Refresh
  % -LRB- CNN -RRB- Kanye West has settled a lawsuit with a paparazzi photographer he assaulted -- and the two have shaken on it.
  % The photographer, Daniel Ramos, had filed the civil suit against West after the hip-hop star attacked him and tried to wrestle his camera from him in July 2013 at Los Angeles International Airport.
  % West pleaded no contest last year to a misdemeanor count of battery over the scuffle.
  
  % PtGen-Covg Output
  % the photographer, daniel ramos, had filed the civil suit against west after the hip-hop star attacked him and tried to wrestle his camera from him in july 2013 at los angeles international airport.
  % west pleaded no contest to a misdemeanor count of battery over the scuffle.
  % ramos and his lawyer, gloria allred, sought general and punitive damages in the civil suit.

  % Gold
  % The rapper assaulted the photographer at Los Angeles International Airport in 2013
  % West apologized as part of the settlement, the photographer 's lawyer says
  \vspace{-0.5cm}
\end{figure}

\paragraph{Baseline BOW Oracle.}

Additionally, we also experimented with a \emph{bag-of-words oracle} (BOW oracle), that serves as a baseline for creating a compressive oracle, in which labels are generated by simply dropping words if they do not appear in the gold summary. Unsurprisingly, oracle sentences compressed with this method are often ungrammatical (see Figure~\ref{fig:oracles-example}). 

Our model \ExConSumm\ trained with the BOW oracle (\ExConSumm~BOW) often score 
higher than the scores of the compressive model as shown in Table \ref{tab:corpus_resultsO}. However, looking at the example summaries in Figure~\ref{fig:summaries-with-BOW}, we find that the BOW compressive model is incapable of generating a fluent or grammatical summary. The \ExConSumm~Compressive summary, on the other hand, is fluent and grammatical. Our summary LSTMs ($\mathrm{WordStates}$ and $\mathrm{SentStates}$) can preserve the fluency of the summaries if trained with the fluent Compressive oracle. This is not guaranteed when using the BOW oracle.

% \subsection{Summary examples}
% \label{sec:appendix2}

% In this section, we provide an example of the summaries generated by the different variants of our model in Fig.~\ref{fig:summaries}. We find that the BoW compressive model, which uses a BoW gold summary for training,  is incapable of generating a fluent or grammatical summary.  The compressive model, however, is able to perform better. 
% This is in line with our hypothesis that the summary LSTMs ($\mathrm{WordEncoder}$ and $\mathrm{SentEncoder}$) can preserve the fluency of the summaries if trained with fluent data.

\subsection{Human Evaluation Experiment Design}
\label{sec:humanQA}
The main assumption behind this evaluation is that the gold summary highlights the most important content of the document. Based on this assumption, the questions are written using the GOLD summary. 
For this study, we used the same 20 documents (10 from CNN and 10 from DailyMail testsets) with an accompanying set of questions based on the gold summary from \newcite{Narayan2018Ranking}.\footnote{The test set for the QA evaluation is publicly available at \url{https://github.com/EdinburghNLP/Refresh}.}
We examined how many questions participants were able to answer by reading system summaries alone, without access to the article. The more questions a system can help to answer, the better it is at summarizing the document as a whole. 
We collected answers from five different participants for each summary and system pair. We marked a correct answer with a score of one, partially correct answers with a score of~0.5, and zero otherwise, the final score is an average of all these scores. Answers were elicited using Amazon's Mechanical Turk crowd-sourcing platform. Examples of systems summaries used for this evaluation are shown in Figures~\ref{fig:summaries-with-qa},~\ref{fig:qa-sample-1},~\ref{fig:qa-sample-2},~\ref{fig:qa-sample-3},~\ref{fig:qa-sample-4},~\ref{fig:qa-sample-5} and~\ref{fig:qa-sample-6}.

\begin{figure*}[t!]
  \center{\footnotesize
    \begin{tabular}{p{15.5cm}}
      \hline \vspace{0.1pt}
        \textbf{\textsc{LEAD}} \\
        \textbullet \hspace{0.1cm} (CNN) Seven people--including Illinois State University associate men's basketball coach Torrey Ward and deputy athletic director Aaron Leetch--died when their small plane crashed while heading back from the NCAA tournament final. \\
        \textbullet \hspace{0.1cm} The aircraft went down overnight Monday about 2 miles east of the Central Illinois Regional Airport in Bloomington, McLean County Sheriff's Office Sgt. Bill Tate said. \\
        \textbullet \hspace{0.1cm} That's about 5 miles from the campus of Illinois State, where Ward and Leetch both worked.
        \\ \vspace{0.1pt}
        \textbf{\textsc{Refresh}} \\
        \textbullet \hspace{0.1cm} (CNN) Seven people--including Illinois State University associate men's basketball coach Torrey Ward and deputy athletic director Aaron Leetch--died when their small plane crashed while heading back from the NCAA tournament final. \\
        \textbullet \hspace{0.1cm} The aircraft went down overnight Monday about 2 miles east of the Central Illinois Regional Airport in Bloomington, McLean County Sheriff's Office Sgt. Bill Tate said. \\
        \textbullet \hspace{0.1cm} The plane was coming back from the NCAA Final Four championship game in Indianapolis, according to Illinois State athletics spokesman John Twork.
        \\ \vspace{0.1pt}
        \textbf{\textsc{Latent}} \\
        \textbullet \hspace{0.1cm} (CNN) Seven people--including Illinois State University associate men's basketball coach Torrey Ward and deputy athletic director Aaron Leetch--died when their small plane crashed while heading back from the NCAA tournament final. \\
        \textbullet \hspace{0.1cm} The plane was coming back from the NCAA Final Four championship game in Indianapolis, according to Illinois State athletics spokesman John Twork. \\
        \textbullet \hspace{0.1cm} The aircraft went down overnight Monday about 2 miles east of the Central Illinois Regional Airport in Bloomington, McLean County Sheriff's Office Sgt. Bill Tate said . 
        \\ \vspace{0.1pt}
        
        \textbf{\ExConSumm\ Extractive} \\
        \textbullet \hspace{0.1cm} (CNN) Seven people--including Illinois State University associate men's basketball coach Torrey Ward and deputy athletic director Aaron Leetch--died when their small plane crashed while heading back from the NCAA tournament final. \\ 
        \textbullet \hspace{0.1cm} The plane was coming back from the NCAA Final Four championship game in Indianapolis, according to Illinois State athletics spokesman John Twork. \\ \hline \vspace{0.1pt}
        
        \textbf{\ExConSumm\ Compressive} \\
        \textbullet \hspace{0.1cm} Seven people died their small plane crashed while heading back from the NCAA tournament final. \\ 
        \textbullet \hspace{0.1cm} The aircraft went down overnight Monday about 2 miles east of the Central Illinois Regional Airport in Bloomington. \\ 
        \textbullet \hspace{0.1cm} The plane was coming back from the NCAA Final Four championship game. \\ \hline   \vspace{0.1pt} 
        \textbf{Pointer+Coverage} \\
        \textbullet \hspace{0.1cm}  Illinois State University associate men's basketball coach Torrey ward and deputy athletic director Aaron Leetch died when their small plane crashed while heading back from the NCAA tournament final. \\
        \textbullet \hspace{0.1cm} The aircraft went down overnight Monday about 2 miles east of the Central Illinois Regional Airport in Bloomington, iIlinois. \\
        \textbullet \hspace{0.1cm} It was not immediately known who else was on the aircraft, which the National Transportation Safety Board tweeted was a Cessna 414. \\
        \textbullet \hspace{0.1cm} There's also a picture of a small plane with the words, ``my ride to the game was n't bad \#indy2015f4''.
        \\ \hline \vspace{0.1pt}

        \textbf{\textsc{GOLD}} \\
        \textbullet \hspace{0.1cm} The crashed plane was a Cessna 414, National Transportation Safety Board reports \\
        \textbullet \hspace{0.1cm} Coach Torrey Ward, administrator Aaron Leetch among the 7 killed in the crash \\
        \textbullet \hspace{0.1cm} The plane crashed while coming back from the NCAA title game in Indianapolis \\ \hline \vspace{0.1pt}
        
        \textbf{Question-Answer Pairs} \\
        \textbullet \hspace{0.1cm}  What type of plane crashed? (Cessna 414) \\
        \textbullet \hspace{0.1cm} Who are confirmed dead in the crash? (Coach Torrey Ward and administrator Aaron Leetch) \\
        \textbullet \hspace{0.1cm} How many people in total died in the crash? (7 people) \\
        \textbullet \hspace{0.1cm} The plane crashed while coming back from where? (The NCAA title game in Indianapolis) \\ \hline
    \end{tabular}
    }
  \caption{ Example output summaries on the CNN/DailyMail dataset, gold standard summary, and corresponding questions.}
  \label{fig:qa-sample-1}
%   Priberam_BOW_Comp
%     Seven people Illinois State University associate men 's basketball coach Torrey Ward and deputy athletic director Aaron died their plane crashed back from the NCAA tournament final 
%     The aircraft went down overnight Monday about 2 miles east of the Central Illinois Regional Airport in
\end{figure*}

\begin{figure*}[t!]
  \center{\footnotesize
    \begin{tabular}{p{16cm}}
      \hline \vspace{0.1pt}
        \textbf{\textsc{LEAD}} \\
        \textbullet \hspace{0.1cm} A hedgehog sniffing around in the dusk was once a common sight - but experts warn it may soon become a thing of the past. \\
        \textbullet \hspace{0.1cm} One in five people have never seen a hedgehog in their gardens, according to a wildlife survey. \\
        \textbullet \hspace{0.1cm} And of those who do spot the tiny animals, only a quarter see them frequently, the RSPB found. \\
        \textbullet \hspace{0.1cm} The startling figures confirm fears that the small British mammal is suffering a huge decline. \\ \vspace{0.1pt}
        \textbf{\textsc{Refresh}} \\
        \textbullet \hspace{0.1cm} One in five people have never seen a hedgehog in their gardens, according to a wildlife survey. \\
        \textbullet \hspace{0.1cm} And of those who do spot the tiny animals, only a quarter see them frequently, the RSPB found. \\
        \textbullet \hspace{0.1cm} The startling figures confirm fears that the small British mammal is suffering a huge decline. \\
        \textbullet \hspace{0.1cm} There are thought to be less than 1 million hedgehogs living in this country today, an estimated 30 per cent drop since 2013. \\ \vspace{0.1pt}
        \textbf{\textsc{Latent}} \\
        \textbullet \hspace{0.1cm} There are thought to be less than 1 million hedgehogs living in this country today, an estimated 30 per cent drop since 2013. \\
        \textbullet \hspace{0.1cm} One in five people have never seen a hedgehog in their gardens, according to a new wildlife survey. \\
        \textbullet \hspace{0.1cm} The startling figures confirm fears that the small British mammal is suffering a huge decline. \\ \vspace{0.1pt}
        
        \textbf{\ExConSumm\ Extractive} \\
        \textbullet \hspace{0.1cm} One in five people have never seen a hedgehog in their gardens, according to a wildlife survey. \\ 
        \textbullet \hspace{0.1cm} And of those who do spot the tiny animals , only a quarter see them frequently, the RSPB found. \\
        \textbullet \hspace{0.1cm} The startling figures confirm fears that the small British mammal is suffering a huge decline. \\
        \textbullet \hspace{0.1cm} There are thought to be less than 1 million hedgehogs living in this country today, an estimated 30 per cent drop since 2013. \\ \hline \vspace{0.1pt}
        
        \textbf{\ExConSumm\ Compressive} \\
        \textbullet \hspace{0.1cm} One in five people have never seen a hedgehog in their gardens. \\ 
        \textbullet \hspace{0.1cm} And of those who do spot the tiny animals, only a quarter see them frequently, the RSPB found. \\
        \textbullet \hspace{0.1cm} The small British mammal is suffering a huge decline. \\ 
        \textbullet \hspace{0.1cm} There are thought to be less than 1 million hedgehogs living in this country today, an estimated 30 per cent drop since 2013. \\ \hline   \vspace{0.1pt}  
      
        \textbf{Pointer+Coverage} \\
        \textbullet \hspace{0.1cm} One in five people have never seen a hedgehog in their gardens. \\
        \textbullet \hspace{0.1cm} Of those who do spot the tiny animals, only a quarter see them frequently. \\
        \textbullet \hspace{0.1cm} The startling figures confirm fears that the small British mammal is suffering a huge decline. \\ \hline \vspace{0.1pt}

        \textbf{\textsc{GOLD}} \\
        \textbullet \hspace{0.1cm} One in five people have never seen a hedgehog in their back gardens \\
        \textbullet \hspace{0.1cm} Only a quarter of those who do admitted seeing the animals frequently \\
        \textbullet \hspace{0.1cm} Wildlife survey suggested the small British mammal is in huge decline \\
        \textbullet \hspace{0.1cm} There are thought to be less than 1 million hedgehogs in the country
        \\ \hline \vspace{0.1pt}
        
        \textbf{Question-Answer Pairs} \\
        \textbullet \hspace{0.1cm}  How many people have never seen a hedgehog in their back gardens? (One in five people) \\
        \textbullet \hspace{0.1cm} Who conducted this survey? (Wildlife survey) \\
        \textbullet \hspace{0.1cm} How many hedgehogs are thought to be left in the country? (less than 1 million) \\ \hline
    \end{tabular}
    }
  \caption{ Example output summaries on the CNN/DailyMail dataset, gold standard summary, and corresponding questions.}
  \label{fig:qa-sample-2}
%   Priberam_BOW_Comp
%     One in five people have never seen a hedgehog in their gardens , to a 
%     of who do spot the tiny animals , only a quarter see them frequently , the RSPB found 
%     that the small British mammal is suffering a huge decline 
%     There are thought to be less than 1 million hedgehogs living in this country today ,
\end{figure*}

\begin{figure*}[t!]
  \center{\footnotesize
    \begin{tabular}{p{16cm}}
      \hline \vspace{0.1pt}
        \textbf{\textsc{LEAD}} \\
        \textbullet \hspace{0.1cm} (CNN) Blinky and Pinky on the Champs Elysees? \\
        \textbullet \hspace{0.1cm} Inky and Clyde running down Broadway? \\
        \textbullet \hspace{0.1cm} Power pellets on the Embarcadero? \\ \vspace{0.1pt}
        \textbf{\textsc{Refresh}} \\
        \textbullet \hspace{0.1cm} Leave it to Google to make April Fools' Day into throwback fun by combining Google Maps with Pac-Man. \\
        \textbullet \hspace{0.1cm} The massive tech company is known for its impish April Fools' Day pranks, and Google Maps has been at the center of a few, including a Pokemon Challenge and a treasure map. \\
        \textbullet \hspace{0.1cm} This year the company was a day early to the party, rolling out the Pac-Man game Tuesday. \\ \vspace{0.1pt}
        \textbf{\textsc{Latent}} \\
        \textbullet \hspace{0.1cm} Leave it to Google to make April Fools' Day into throwback fun by combining Google Maps with Pac-Man. \\
        \textbullet \hspace{0.1cm} (CNN) Blinky and Pinky on the Champs Elysees? Inky and Clyde running down Broadway? Power pellets on the Embarcadero? \\
        \textbullet \hspace{0.1cm} Twitterers have been tickled by the possibilities, playing Pac-Man in Manhattan, on the University of Illinois quad, in central London and down crooked Lombard Street in San Fran cisco, among many locations:. \\ \vspace{0.1pt}
        
        \textbf{\ExConSumm\ Extractive} \\
        \textbullet \hspace{0.1cm} Leave it to Google to make April Fools' Day into throwback fun by combining Google Maps with Pac-Man. \\ 
        \textbullet \hspace{0.1cm} It's easy to play: Simply pull up Google Maps on your desktop browser, click on the Pac-Man icon on the lower left, and your map suddenly becomes a Pac-Man course. \\ \hline \vspace{0.1pt}
        
        \textbf{\ExConSumm\ Compressive} \\
        \textbullet \hspace{0.1cm}  Leave it to Google to make April Fools Day into throwback fun by combining Google Maps with Pac-Man. \\ 
        \textbullet \hspace{0.1cm} The tech company is known for its April Fools Day pranks. \\ \hline   \vspace{0.1pt}  
      
        \textbf{Pointer+Coverage} \\
        \textbullet \hspace{0.1cm} The massive tech company is known for its impish April fools' day pranks, and Google Maps has been at the center of a few, including a Pokemon challenge and a treasure map. \\
        \textbullet \hspace{0.1cm}  It's easy to play : simply pull up Google Maps on your desktop browser. \\ \hline \vspace{0.1pt}

        \textbf{\textsc{GOLD}} \\
        \textbullet \hspace{0.1cm}  Google Maps has a temporary Pac-Man function \\
        \textbullet \hspace{0.1cm} Google has long been fond of April Fools' Day pranks and games \\
        \textbullet \hspace{0.1cm} Many people are turning their cities into Pac-Man courses \\ \hline \vspace{0.1pt}
        
        \textbf{Question-Answer Pairs} \\
        \textbullet \hspace{0.1cm} What function does Google Maps have? (Pac-Man) \\
        \textbullet \hspace{0.1cm} What has Google been long fond of? (April Fools' Day pranks and games) \\
        \textbullet \hspace{0.1cm} What are many people turning their cities into? (Pac-Man courses)
 \\ \hline
    \end{tabular}
    }
  \caption{ Example output summaries on the CNN/DailyMail dataset, gold standard summary, and corresponding questions.}
  \label{fig:qa-sample-3}
%   Priberam_BOW_Comp
% Power pellets on the 
% to Google to make April Fools ' Day into throwback fun by Google Maps
\end{figure*}

\begin{figure*}[t!]
  \center{\footnotesize
    \begin{tabular}{p{16cm}}
      \hline \vspace{0.1pt}
        \textbf{\textsc{LEAD}} \\
        \textbullet \hspace{0.1cm} (CNN) Somewhere over the rainbow, people on the Internet are losing their minds. \\
        \textbullet \hspace{0.1cm} Is it real? \\
        \textbullet \hspace{0.1cm} After the New York area received a large amount of rain, four rainbows stretched across the early morning sky on Tuesday. \\ \vspace{0.1pt}
        \textbf{\textsc{Refresh}} \\
        \textbullet \hspace{0.1cm} (CNN) Somewhere over the rainbow, people on the Internet are losing their minds. \\
        \textbullet \hspace{0.1cm} After the New York area received a large amount of rain, four rainbows stretched across the early morning sky on Tuesday. \\
        \textbullet \hspace{0.1cm} Amanda Curtis, CEO of a fashion company in New York, snapped the lucky shot. \\ \vspace{0.1pt}
        \textbf{\textsc{Latent}} \\
        \textbullet \hspace{0.1cm} Amanda Curtis, CEO of a fashion company in New York, snapped the lucky shot. \\
        \textbullet \hspace{0.1cm} After the New York area received a large amount of rain, four rainbows stretched across the early morning sky on Tuesday. \\
        \textbullet \hspace{0.1cm} CNN iReporter Yosemitebear Vasquez posted a video to YouTube in 2010 reacting to a double rainbow he spotted in Yosemite National Park. The video has since garnered over 40 million views. \\ \vspace{0.1pt}
        
        \textbf{\ExConSumm\ Extractive} \\
        \textbullet \hspace{0.1cm} After the New York area received a large amount of rain, four rainbows stretched across the early morning sky on Tuesday. \\ 
        \textbullet \hspace{0.1cm} Amanda Curtis, CEO of a fashion company in New York, snapped the lucky shot. \\
        \textbullet \hspace{0.1cm} The video has since garnered over 40 million views. \\ \hline \vspace{0.1pt}
        
        \textbf{\ExConSumm\ Compressive} \\
        \textbullet \hspace{0.1cm} Four rainbows stretched across the early morning sky on Tuesday. \\ 
        \textbullet \hspace{0.1cm} Amanda Curtis, CEO of a fashion company in New York, snapped the lucky shot. \\
        \textbullet \hspace{0.1cm} The video has since garnered over 40 million views. \\ \hline   \vspace{0.1pt}  
      
        \textbf{Pointer+Coverage} \\
        \textbullet \hspace{0.1cm} Amanda Curtis, CEO of a fashion company in New York, snapped the lucky shot. \\
        \textbullet \hspace{0.1cm} She posted the picture to Twitter, and within a few hours, it had already received hundreds of retweets. \\ \hline \vspace{0.1pt}

        \textbf{\textsc{GOLD}} \\
        \textbullet \hspace{0.1cm} Amanda Curtis, CEO of a fashion company in New York, posted a picture of four rainbows to Twitter \\
        \textbullet \hspace{0.1cm} ``I had a small moment of awe,'' she said  \\ \hline \vspace{0.1pt}
        
        \textbf{Question-Answer Pairs} \\
        \textbullet \hspace{0.1cm}  Who posted a picture to Twitter? (Amanda Curtis) \\
        \textbullet \hspace{0.1cm} What did the picture show? (four rainbows) \\
        \textbullet \hspace{0.1cm} What is the profession of the person who posted this picture? (CEO of a fashion company in New York) \\ \hline
    \end{tabular}
    }
  \caption{ Example output summaries on the CNN/DailyMail dataset, gold standard summary, and corresponding questions.}
  \label{fig:qa-sample-4}
%   Priberam_BOW_Comp
% the New York area a large amount of rain , four rainbows stretched across the early morning sky on Tuesday 
% Amanda Curtis , CEO of a fashion company in New York , the
\end{figure*}

\begin{figure*}[t!]
  \center{\footnotesize
    \begin{tabular}{p{16cm}}
      \hline \vspace{0.1pt}
        \textbf{\textsc{LEAD}} \\
        \textbullet \hspace{0.1cm} The Fulham fans in the Jimmy Steed Stand applauded their team at the final whistle. \\
        \textbullet \hspace{0.1cm} It was not the victory manager Kit Symons had called for but a point away to Charlton probably secures their future in the Championship next season. \\
        \textbullet \hspace{0.1cm} They are eight points clear of Millwall and bar a miraculous resurgence from one of the bottom three sides will stay up but the fact relegation is still mathematically feasible for a club that were in the Premier League last season is alarming. \\
        \textbullet \hspace{0.1cm} A vertiginous decline, just one victory in their last seven games had seen them dragged back into a relegation battle and after a painful 4-1 trouncing by bitter rivals Brentford last week, Symons was looking for a quick response from his players. \\ \vspace{0.1pt}
        \textbf{\textsc{Refresh}} \\
        \textbullet \hspace{0.1cm} The Fulham fans in the Jimmy Steed Stand applauded their team at the final whistle. \\
        \textbullet \hspace{0.1cm} It was not the victory manager Kit Symons had called for but a point away to Charlton probably secures their future in the Championship next season. \\
        \textbullet \hspace{0.1cm} They are eight points clear of Millwall and bar a miraculous resurgence from one of the bottom three sides will stay up but the fact relegation is still mathematically feasible for a club that were in the Premier League last season is alarming. \\
        \textbullet \hspace{0.1cm} He got it with Ross McCormack giving them the lead after eight minutes. \\ \vspace{0.1pt}
        \textbf{\textsc{Latent}} \\
        \textbullet \hspace{0.1cm} Johann Gudmundsson celebrates his first-half effort as Charlton come from behind to earn a point. \\
        \textbullet \hspace{0.1cm} Ross McCormack headed over a stranded Stephen Henderson with just eight minutes played in London. \\
        \textbullet \hspace{0.1cm} Fulham now sit 20th in the table eight points clear of fellow London rivals Millwall, but Symons is refusing to relax just yet. \\ \vspace{0.1pt}
        
        \textbf{\ExConSumm\ Extractive} \\
        \textbullet \hspace{0.1cm} The Fulham fans in the Jimmy Steed Stand applauded their team at the final whistle. \\ 
        \textbullet \hspace{0.1cm} It was not the victory manager Kit Symons had called for but a point away to Charlton probably secures their future in the Championship next season . \\ \hline \vspace{0.1pt}
        
        \textbf{\ExConSumm\ Compressive} \\
        \textbullet \hspace{0.1cm}  The Fulham fans in the Jimmy Steed Stand applauded their team at the final whistle. \\ 
        \textbullet \hspace{0.1cm} It was not the victory manager Kit Symons had called for but a point away. \\ 
        \textbullet \hspace{0.1cm} They are eight points clear of Millwall and bar a miraculous resurgence. \\ 
        \textbullet \hspace{0.1cm} Fulham sit 20th in the table eight points clear of fellow London rivals, but Symons is refusing to relax just yet. \\ \hline   \vspace{0.1pt}  
      
        \textbf{Pointer+Coverage} \\
        \textbullet \hspace{0.1cm} Fulham fans in the Jimmy Steed Stand applauded their team at the final whistle. \\
        \textbullet \hspace{0.1cm} It was not the victory manager Kit Symons had called for but a point away to Charlton probably secures their future in the Championship next season. \\
        \textbullet \hspace{0.1cm} They are eight points clear of Millwall and bar a miraculous resurgence. \\ \hline \vspace{0.1pt}

        \textbf{\textsc{GOLD}} \\
        \textbullet \hspace{0.1cm} Ross McCormack gave Fulham the lead after eight minutes at The Valley \\
        \textbullet \hspace{0.1cm} But Johann Gudmundsson leveled the scores less than ten minutes later \\
        \textbullet \hspace{0.1cm} Scott Parker was booed on his return to club, 11 years after he left \\
        \textbullet \hspace{0.1cm} Share of the points in London leaves Charlton in 11th and Fulham in 20th \\ \hline \vspace{0.1pt}
        
        \textbf{Question-Answer Pairs} \\
        \textbullet \hspace{0.1cm} Who gave Fulham the lead after eight minutes at The Valley? (Ross McCormack) \\
        \textbullet \hspace{0.1cm} Who leveled the scores less than ten minutes later? (Johann Gudmundsson) \\
        \textbullet \hspace{0.1cm} Who was booed on his return to club, 11 years after he left? (Scott Parker) \\
        \textbullet \hspace{0.1cm} What is Charlton's place after the share of the points in London? (11th) \\
        \textbullet \hspace{0.1cm} What is Fulham's place after the share of the points in London? (20th) \\ \hline
    \end{tabular}
    }
  \caption{ Example output summaries on the CNN/DailyMail dataset, gold standard summary, and corresponding questions.}
  \label{fig:qa-sample-5}
%   Priberam_BOW_Comp
% The Fulham fans in the Jimmy Steed Stand applauded their team at the final whistle 
% was not the victory manager Kit Symons had called for but a to Charlton in the
\end{figure*}

\begin{figure*}[t!]
  \center{\footnotesize
    \begin{tabular}{p{16cm}}
      \hline \vspace{0.1pt}
        \textbf{\textsc{LEAD}} \\
        \textbullet \hspace{0.1cm} (CNN) You probably never knew her name, but you were familiar with her work. \\
        \textbullet \hspace{0.1cm} Betty Whitehead Willis, the designer of the iconic ``Welcome to Fabulous Las Vegas'' sign, died over the weekend. \\
        \textbullet \hspace{0.1cm} She was 91. \\ \vspace{0.1pt}
        \textbf{\textsc{Refresh}} \\
        \textbullet \hspace{0.1cm} Betty Whitehead Willis, the designer of the iconic ``Welcome to Fabulous Las Vegas'' sign, died over the weekend. \\
        \textbullet \hspace{0.1cm} Willis played a major role in creating some of the most memorable neon work in the city. \\
        \textbullet \hspace{0.1cm} Willis visited the Neon Museum in 2013 to celebrate her 90th birthday. \\ \vspace{0.1pt}
        \textbf{\textsc{Latent}} \\
        \textbullet \hspace{0.1cm} Betty Whitehead Willis, the designer of the iconic ``Welcome to Fabulous Las Vegas'' sign, died over the weekend. She was 91. \\
        \textbullet \hspace{0.1cm} The Neon Museum also credits her with designing the signs for Moulin Rouge Hotel and Blue Angel Motel. \\
        \textbullet \hspace{0.1cm} Willis visited the Neon Museum in 2013 to celebrate her 90th birthday. \\ \vspace{0.1pt}
        
        \textbf{\ExConSumm\ Extractive} \\
        \textbullet \hspace{0.1cm} Betty Whitehead Willis, the designer of the iconic ``Welcome to Fabulous Las Vegas'' sign, died over the weekend. \\ 
        \textbullet \hspace{0.1cm} Willis visited the Neon Museum in 2013 to celebrate her 90th birthday. \\ \hline \vspace{0.1pt}
        
        \textbf{\ExConSumm\ Compressive} \\
        \textbullet \hspace{0.1cm} Betty Whitehead Willis died over the weekend. \\ 
        \textbullet \hspace{0.1cm} Willis played a major role in creating some of the most memorable neon work in the city. \\
        \textbullet \hspace{0.1cm} Willis visited the Neon Museum in 2013 to celebrate her 90th birthday.  \\ \hline   \vspace{0.1pt}  
      
        \textbf{Pointer+Coverage} \\
        \textbullet \hspace{0.1cm} Betty Whitehead Willis, the designer of the iconic ``Welcome to Fabulous Las Vegas, died over the weekend. \\
        \textbullet \hspace{0.1cm} She was 91. \\
        \textbullet \hspace{0.1cm} Willis never trademarked her most-famous work, calling it ``my gift to the city''. \\ \hline \vspace{0.1pt}

        \textbf{\textsc{GOLD}} \\
        \textbullet \hspace{0.1cm} Willis never trademarked her most-famous work, calling it ``my gift to the city'' \\
        \textbullet \hspace{0.1cm} She created some of the city's most famous neon work  \\ \hline \vspace{0.1pt}
        
        \textbf{Question-Answer Pairs} \\
        \textbullet \hspace{0.1cm} What was Willis' most-famous work called? (my gift to the city) \\
        \textbullet \hspace{0.1cm} What did Willis create in the city? (City's most famous neon work) \\ \hline
    \end{tabular}
    }
  \caption{ Example output summaries on the CNN/DailyMail dataset, gold standard summary, and corresponding questions.}
  \label{fig:qa-sample-6}
%   Priberam_BOW_Comp
% Betty Whitehead Willis , the designer of the Welcome to Fabulous Las Vegas , died the 
% Willis played a major role in creating some of the most memorable neon work in the
\end{figure*}

\end{document}